\documentclass[conference]{IEEEtran}

\usepackage{graphicx} \usepackage{balance} \usepackage{hyperref} \usepackage[all]{nowidow} \usepackage{float}
\usepackage{tikz} \usepackage{tikz-3dplot} \usepackage{nicefrac} \usepackage{algorithm} \usepackage{subcaption}
\usepackage[noend]{algpseudocode} \usepackage{array} \usepackage{pgfplots} \usepackage{booktabs} \usepackage{siunitx}
\usepackage{enumitem} \usepackage[utf8]{inputenc} \usepackage{amsmath} \usepackage{amssymb} \usepackage{filecontents}

\begin{document}
    \title{Empirical Analysis of Common Subgraph Isomorphism Approaches to the Lost-in-Space Star
    Identification Problem}

    \author{
    \IEEEauthorblockN{Glenn Galvizo}
    \IEEEauthorblockA{
    University of Hawaii at Manoa \\
    glennga@hawaii.edu
    }
    \and
    \IEEEauthorblockN{Lipyeow Lim}
    \IEEEauthorblockA{
    University of Hawaii at Manoa \\
    lipyeow@hawaii.edu
    }
    }

    \maketitle

    \begin{abstract}
        The process of identifying stars is integral toward stellar based orientation determination in spacecraft.
        Star identification involves matching points in an image of the sky with stars in an astronomical catalog.
        A unified framework for identification was created and used to analyze six variations of methods based on their
        approach to star set identification, obtaining a single image to catalog star set match, and uniquely mapping
        each star in a image star set to a catalog star set.
        Each method was presented an artificial image, and aspects that were interchangeable among each process were
        normalized.
        Given an image with false stars, the Pyramid method has the highest average accuracy and is the fastest of the six.
        Given an image where each star's true position is distributed randomly (Gaussian noise), the Spherical Triangle
        method's accuracy is the least sensitive.
    \end{abstract}

    \section{Introduction}\label{sec:introduction}
    Ancient mariners could look up at the night sky, point out which stars they were looking at, and navigate across the
    globe without the use of maps.
    \textit{Star identification algorithms} refer to computational approaches to determining which stars are in the sky.
    Given an image of the sky, star identification is matching the bright spots in an image to stars in an astronomical
    catalog.
    The device that performs these computations is the star tracker, much like the navigators on the ship.
    \textit{Lost-in-space} refers to an additional constraint on the problem: the absence of knowing where we took
    the picture and how we pointed the camera.

    This problem is most prevalent in designing LEO (low Earth orbit) spacecraft.
    In order for a craft to point a payload, direct its thrusters, or orient its solar panels, an accurate
    \textit{attitude} (another term for orientation) must be known.
    There are a few known landmarks in space where some attitude can be extracted (the Earth, the Sun), but this
    requires constant direction towards just these objects.
    Star trackers do not limit themselves to a single object, rather they use multiple stars within their field of view
    to determine their orientation.

    \newcommand{\seq}{\!=\!}
    There exist roughly $4{,}500$ stars in the sky visible to the human eye.
    For an image of $n$ stars, the naive approach would be compute $C(4{,}500, n)$ combinations from this collection and
    compare each to some subset of stars found in the image.
    For $n\seq 3$, this requires over $10^{10}$ comparisons.
    As an alternative, we sacrifice storage and precision for speed by searching a separate collection which indexes the
    ${\sim}4{,}500$ stars by one or more features.
    When this subset is identified, we determine and return the orientation of the image relative to collection
    of ${\sim}4{,}500$ stars.

    This research is motivated by a growing difference in the number of stellar attitude determination methods and
    empirical comparison between each of these methods in a more systematic manner for star tracker development.
    Interchangeable factors are abstracted away (camera hardware, blob detection, etc\ldots) to focus more on how each
    method matches stars in an image to stars in a catalog.
    This paper aims to contribute a hardware independent comparison process, an algorithmic description of several
    identification methods, as well as runtime and catalog access analysis of these methods under various types of noise.
    The process of identifying blobs in an image, constructing the image coordinate system, and efficiently querying
    static databases are not addressed here.

    \newcommand{\iFrame}{\mathcal{I}}
    \newcommand{\kFrame}{\mathcal{K}}
    \newcommand{\vv}[1]{#1} 

    \section{Attitude Determination}\label{sec:attitudeDetermination}

    This section serves to give a brief overview into what attitude is, what Wahba's problem is \& how to solve it, and what
    stellar based attitude determination entails.

    \subsection{General Attitude Determination}\label{subsec:generalAttitudeDetermination}
    Attitude refers to the translation between how one system describes an object compared to how a different system
    describes the same object.
    These systems are referred to as \textit{reference frames}, and describe objects in terms of dimensions ($x_1, x_2, x_3,
    \ldots$).
    As an example, observer $A$ at the bottom of a mountain may describe the mountain itself as large and above itself.
    Another observer $B$ on a helicopter hovering over the same mountain may describe it as small and below itself.
    To find an attitude between $A$'s reference frame and $B$'s reference frame is to find some function $h(x_1, x_2, x_3,
    \ldots)$ that is able to produce $B$'s description of the mountain with $A$'s observations.

    In the context of spacecraft attitude from star identification, there exist three reference frames: the
    \textit{body frame}, the \textit{sensor frame}, and the \textit{inertial frame}.
    The body frame itself is fixed to the structure of the spacecraft, the sensor frame is fixed to the star tracker,
    and the inertial frame refers to some non-accelerating frame in which stellar objects are recorded.
    All observations from the spacecraft exist in the sensor frame, but can easily be rotated to align with the body frame
    (the sensor itself is fixed to the spacecraft chassis).
    Consequently, the body frame is used interchangeably with the sensor frame.
    To describe the craft itself, an inertial frame is required for finding a practical attitude.
    A star observed in the inertial frame is more predictable than the same star observed in a tumbling spacecraft, aiding
    the usage of the attitude with orientation dependent processes.
    Using all three, the goal of attitude determination becomes finding some method of translation between the inertial
    frame and the body frame.

    \begin{figure}
        \centering{
        \tdplotsetmaincoords{60}{110}

        \pgfmathsetmacro{\rvec}{.8} \pgfmathsetmacro{\thetavec}{30} \pgfmathsetmacro{\phivec}{60}

        \begin{tikzpicture}[scale=3,tdplot_main_coords]

            \coordinate (O) at (0,0,0);
            \node[color=red, anchor=east] at (0,0,0) {$\kFrame$ Frame};
            \draw[thick,->] (0,0,0) -- (0.7,0,0) node[anchor=north east]{$\vv{u_1}$};
            \draw[thick,->] (0,0,0) -- (0,0.7,0) node[anchor=north west]{$\vv{u_2}$};
            \draw[thick,->] (0,0,0) -- (0,0,0.7) node[anchor=south]{$\vv{u_3}$};

            \tdplotsetcoord{P}{\rvec}{\thetavec}{\phivec}

            \tdplotsetrotatedcoords{\phivec}{\thetavec}{0}
            \tdplotsetrotatedcoordsorigin{(P)}

            \draw[tdplot_rotated_coords] (.205,.205,.205) circle[radius=0.3pt,fill=gray];
            \node[tdplot_rotated_coords, anchor=west, xshift=1cm, color=blue] at (.2,.2,.2) {$\vv{I_j}$};

            \draw[-stealth,color=blue,tdplot_rotated_coords] (0,0,0) -- (.2,.2,.2);
            \draw[dashed,color=blue,tdplot_rotated_coords] (0,0,0) -- (.2,.2,0);
            \draw[dashed,color=blue,tdplot_rotated_coords] (.2,.2,0) -- (.2,.2,.2);

            \coordinate (Q) at (0.4,0.767,0.91);
            \draw[-stealth,color=red] (O) -- (Q);
            \node[color=red,fill=white,xshift=-0.1cm,yshift=-0.1cm] at (0.2,0.3835,0.455) {$\vv{K_j}$};
            \draw[dashed, color=red] (O) -- (0.4,0.767,0);
            \draw[dashed, color=red] (Q) -- (0.4,0.767,0);

            \node[color=blue,tdplot_rotated_coords,anchor=east,xshift=1.1cm, yshift=1.2cm] at (0,0,0) {$\iFrame$ Frame};
            \draw[thick,tdplot_rotated_coords,->] (0,0,0) -- (.5,0,0) node[anchor=east]{$\vv{v_1}$};
            \draw[thick,tdplot_rotated_coords,->] (0,0,0) -- (0,.5,0) node[anchor=west]{$\vv{v_2}$};
            \draw[thick,tdplot_rotated_coords,->] (0,0,0) -- (0,0,.5) node[anchor=west]{$\vv{v_3}$};

        \end{tikzpicture}

        \caption{
        Visual of two coordinate frames: the inertial frame $\kFrame$, and the body frame $\iFrame$.
        Observation $j$ is described with vector $\vv{I_j}$ in the $\iFrame$ body frame.
        The same observation $j$ is described with vector $\vv{K_j}$ in the $\kFrame$ inertial frame.
        By aligning several observations in both frames, a spacecraft orientation $A^{\nicefrac{\iFrame}{\kFrame}}$
        can be determined to take all points in $\kFrame$ to $\iFrame$.
        }
        \label{figure:coordinateSystem}
        }
    \end{figure}
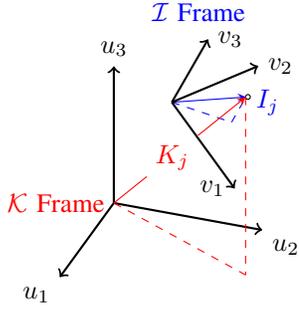

    \begin{subequations}
        ~\autoref{figure:coordinateSystem} describes an inertial frame $\kFrame$ with a right hand set of three orthogonal
        vectors $\{\vv{u_1}, \vv{u_2}, \vv{u_3}\}$ and a body frame $\iFrame$ with another right hand set of three
        orthogonal vectors $\{\vv{v_1}, \vv{v_2}, \vv{v_3}\}$~\cite{wie:spaceVehicleDynamics}.
        A \textit{rotation matrix} $A$ can be assembled to describe the basis vectors of $\kFrame$ in terms of $\iFrame$:
        \begin{align}
            \begin{bmatrix}
                \vv{v_1} \\
                \vv{v_2}  \\
                \vv{v_3}
            \end{bmatrix} &=
            \begin{bmatrix}
                \vv{v_1} \cdot \vv{u_1} & \vv{v_1} \cdot \vv{u_2} & \vv{v_1} \cdot \vv{u_3} \\
                \vv{v_2} \cdot \vv{u_1} & \vv{v_2} \cdot \vv{u_2} & \vv{v_2} \cdot \vv{u_3} \\
                \vv{v_3} \cdot \vv{u_1} & \vv{v_3} \cdot \vv{u_2} & \vv{v_3} \cdot \vv{u_3}
            \end{bmatrix}
            \begin{bmatrix}
                \vv{u_1} \\
                \vv{u_2}  \\
                \vv{u_3}
            \end{bmatrix} \\
            \begin{bmatrix}
                \vv{v_1} \\
                \vv{v_2}  \\
                \vv{v_3}
            \end{bmatrix} &=
            A^{\nicefrac{\iFrame}{\kFrame}}
            \begin{bmatrix}
                \vv{u_1} \\
                \vv{u_2}  \\
                \vv{u_3}
            \end{bmatrix}
        \end{align}
    \end{subequations}
    The issue here is that this rotation matrix $A$ is not given and that we need to account for the noise associated with
    our measurements.
    This problem is known as \textit{Wahba's problem}, first posed by Gracie Wahba in
    1965~\cite{wahba:attitudeEstimationProblem}.
    Wahba's problem states that finding the optimal $A$ involves minimizing the loss function below:
    \begin{equation}
        L(A) = \frac{1}{2} \sum_j^n \vv{w_j} \left\| \vv{I_j} - A\vv{K_j} \right\|^2
    \end{equation}
    where $\vv{w_j}$ represents a non negative weight associated with the noise between the observations $\vv{I_j}$
    in the body frame and $\vv{K_j}$ in the inertial frame.

    \begin{subequations}
        For $n \!>\! 2$, Wahba's problem exists as an optimization problem.
        In the $n\seq2$ case though, the \textit{TRIAD method} (short for TRIaxial Attitude Determination) exists as a
        closed form solution~\cite{markley:attitudeDeterminationTwoVectors}.
        This algorithm starts by constructing two sets of basis vectors: one attached to the body referential (two
        observations in the body frame) $\left[ \vv{t_{1I}} \ \vv{t_{2I}} \ \vv{t_{3I}} \right]$ and another attached to
        the inertial referential (two observations in the inertial frame) $\left[ \vv{t_{2I}} \ \vv{t_{2K}} \ \vv{t_{3K}}
        \right]$~\cite{benet:swisscubeAttitudeDetermination,black:passiveAttitudeDetermination}.
        This is known as the triad frame:
        \begin{alignat}{4}
            \vv{t_{1I}} &= \frac{\vv{v_1}}{\left| \vv{v_1} \right|} &\vv{t_{2I}} &{}={}&
            \frac{\vv{u_1}}{\left| \vv{u_1} \right|} \ \ \ \ \ \ \  \\
            \vv{t_{2I}} &= \frac{\vv{v_1} \times \vv{v_2}}{\left| \vv{v_1} \times \vv{v_2} \right|} \ \ \ \ \ \ \ \
            &\vv{t_{2K}} &{}={}& \frac{\vv{u_1} \times \vv{u_2}}{\left| \vv{u_1} \times \vv{u_2} \right|} \\
            \vv{t_{3I}} &= \vv{t_{1I}} \times \vv{t_{2I}} &\vv{t_{3K}} &{}={}& \vv{t_{2I}} \times \vv{t_{2K}}
        \end{alignat}
    \end{subequations}
    Getting from frame $\kFrame$ to $\iFrame$ now simplifies to multiplication of the triad frame base change matrices:
    \begin{equation}
        A =
        \begin{bmatrix}
            \vv{t_{1K}} & \vv{t_{2K}} & \vv{t_{3K}}
        \end{bmatrix}
        \begin{bmatrix}
            \vv{t_{1I}} & \vv{t_{2I}} & \vv{t_{3I}}
        \end{bmatrix}^T
    \end{equation}
    For all instances where a rotation between the inertial and body frames was required, the TRIAD algorithm was used.

    \subsection{Stellar Based Attitude Determination}\label{subsec:stellarBasedAttitudeDetermination}
    \begin{subequations}
        Relative to our solar system, the majority of bright stars ($m \!<\! 6.0$, or visible from the Earth with the naked
        eye) do not visibly move.
        For simplicity, we make the assumption here that all stars in $\kFrame$ are fixed and exist in a inertial frame
        known as the \textit{Earth centered inertial} frame, or ECI frame.
        The star vectors themselves come from star catalogs, the majority of which use the ECI frame and record the
        positions of stars as points lying on a sphere known as the celestial sphere~\cite{tappe:starTrackerDevelopment}.
        Two pieces of information are given here: right ascension $\alpha$ (equivalent to latitude on Earth) and
        declination $\delta$ (equivalent to longitude).
        Representing some spherical point $(\alpha, \delta, r)$ in 3D Cartesian space involves the following:
        \begin{align} \label{eq:sphereToCartesian}
        x &= r \cos(\delta) \cos(\alpha) \\
        y &= r \cos(\delta) \sin(\alpha) \\
        z &= r \sin(\delta)
        \end{align}
        where both $\alpha$ and $\delta$ are in degrees, and $r$ represents some constant distance from Earth.
        $\vv{K_j}$ represents a point obtained from a star catalog that lies in the ECI frame, $r$ units away from Earth.
    \end{subequations}

    Let $\vv{I_j}$ represent a 3D point projected from a 2D observation taken by the star tracker.
    A basic star tracker is composed of a camera, a computer for determining orientation, and a link back to the main
    computer.
    After taking the picture, the pixel positions of potential stars in the image are determined.
    This involves finding bright blobs in the image, and computing each blob's center of mass to get a point ($x, y$).
    Through some 2D to 3D transformation process involving the camera's hardware (i.e.\ field of view, focal point,
    etc\ldots), a 3D point is then obtained~\cite{tappe:starTrackerDevelopment}.

    The next issue is the focus of this paper: determining which observation from the star tracker frame $\iFrame$
    corresponds to which observation from the star catalog frame $\kFrame$.
    Once this correspondence is found, Wahba's problem is solved to obtain $A$ and this is returned to the main computer.

    \newcommand{\nsubparagraph}[1]{\textbf{#1}}

    \section{Related Work}\label{sec:relatedWork}
    This section serves to give a brief overview into the different approaches to the lost-in-space star identification
    problem.
    More comprehensive survey papers have been published by Spratling~\cite{spratling:surveyStarIdentification} and
    Br\"{a}tt~\cite{bratt:analysisStarIdentification}.

    \nsubparagraph{Identification Classes:}
    The first main class of identification and the focus of this paper is the \textit{subgraph isomorphism} class.
    Subgraph isomorphism is NP complete problem which aims to find some 1-to-1 mapping between the vertices (stars) in two
    graphs (i.e.\ the catalog and the image) if it exists~\cite{scott:graphIsomorphismProblem}.
    This involves describing and mapping sets of stars between both the catalog and image in terms of their features
    relative to each other.

    The second class of identification is the \textit{pattern recognition} class.
    In contrast to subgraph isomorphism class, the pattern recognition class commonly deals with larger star sets within
    some defined field-of-view and matches patterns rather than features.
    Pattern formation typically involves 2D binary matrices (grids), where `1' occupies a cell with a star and `0' occupies
    a cell without one~\cite{padgett:gridAlgorithm}.

    %
    %
    %

    \nsubparagraph{Recursive Property:}
    Recall that the lost-in-space condition specifies that we do not have any information about the spacecraft's attitude
    prior to starting our identification algorithm.
    For the majority of a star tracker's lifetime though, this constraint can be relaxed to allow for the use of
    \textit{recursive} star identification.
    Recursive methods possess an attitude recorded at time $t$, and perform the identification at a later time $t + dt$.
    Two methods proposed by Samaan (SP-Search and SNA) reduce the amount of candidate stars from the catalog that could
    map to stars from the image~\cite{samaan:recursiveMode}.

    \nsubparagraph{Features:}
    Each star has a position associated with it, be it from a star catalog or from the image.
    Using this position, the most common feature is the interstar angle between two stars, first utilized by Gottlieb to
    identify sets of three stars with three angles~\cite{gottlieb:spacecraftAttitudeDetermination}.
    Notable methods with geometric functions utilizing these interstar angles were proposed by:
    Groth~\cite{groth:patternMatchingMethod}, Cole \&
    Crassidus~\cite{coleAndCrassidis:sphericalTriangleMethod,coleAndCrassidis:planarTriangleMethod}, and
    Lang~\cite{lang:astrometryDotNet}.
    Another common feature is the interior angle between three stars, where one star exists as a vertex to two other stars.
    Liebe uses this in conjunction with interstar angles~\cite{liebe:starTrackersAttitudeDetermination}.



    Each star also has a brightness attached it, a feature less commonly used due to large variance in measurement.
    Spratling describes two early methods to take advantage of this feature.
    Scholl proposed the usage of this to remove the need for ambiguity after matching star subsets with angular features
    ~\cite{scholl:starFieldIdentification}.
    Ketchum later introduced the second sequential filtering algorithm, which identifies two stars using their brightness
    in comparison to the common trio required of interstar angle methods~\cite{ketchum:onboardStarIdentification}.
    More recent work toward integrating brightness more heavily has been performed by Zhang et
    al~\cite{zhang:brightnessReferenced}.

    \newcommand{\srightarrow}{\! \rightarrow \!}
    \begin{figure}[ht]
        \centering{
        \usetikzlibrary{shapes.geometric, arrows}

        \tikzstyle{process} = [rectangle, text width=2cm, minimum width=2cm, minimum height=0.5cm,text centered, draw=black,
        fill=orange!30]

        \tikzstyle{terminal} = [rectangle, text width=2cm, minimum width=2cm, minimum height=0.5cm,text centered,
        draw=black, fill=red!30]

        \tikzstyle{decision} = [diamond, text width=1.5cm, minimum width=2.2cm, minimum height=1.8cm,text centered, draw=black,
        fill=green!30, inner sep=-12pt]

        \tikzstyle{line} = [draw, -latex']

        \begin{tikzpicture}[node distance=1.2cm]
            \node[scale=1](getImage)[terminal]{Get Camera Image};
            \node[scale=1](pickQueryStars) [process, left of=getImage, xshift = -2.2cm] {Pick $d$ Image Stars};
            \node[scale=1](searchCatalog)[process, below of=pickQueryStars, yshift=-0.4cm] {Query Catalog};
            \node[scale=1](confidentInCatalog)[decision, below of=searchCatalog, yshift=-0.4cm] {$\lvert R \rvert > 0$?};
            \node[scale=1](filterCandidates)[process, below of=confidentInCatalog, yshift=-0.4cm] {Select Candidate};
            \node[scale=1](confidentAfterFilter)[decision, below of=filterCandidates, yshift=-0.4cm] {Confident?};
            \node[scale=1](findMap)[process, below of=confidentAfterFilter, yshift=-0.4cm]{Identify};
            \node[scale=1](confidentInMap)[decision, below of=findMap, yshift=-0.4cm] {Confident?};
            \node[scale=1](returnMap)[terminal, right of=confidentInMap, xshift = 2.2cm] {Return $b, r, h$};

            \draw[->,>=stealth](getImage) -- node[scale=1.3, yshift=-0.3cm]{$I$}(pickQueryStars);
            \draw[->,>=stealth] (pickQueryStars) -- node[scale=1.3, xshift=0.5cm]{$b$}(searchCatalog);
            \draw[->, >=stealth] (searchCatalog) -- node[scale=1.3, xshift=0.5cm, yshift=-0.15cm]{$R$}(confidentInCatalog);
            \draw[->, >=stealth] (confidentInCatalog) -- node[anchor=east, yshift=0.1cm]{Yes}(filterCandidates);
            \draw[->, >=stealth] (filterCandidates) --
            node[scale=1.3, xshift=0.5cm, yshift=-0.15cm]{$r$}(confidentAfterFilter);
            \draw[->, >=stealth] (confidentAfterFilter) -- node[anchor=east, yshift=0.1cm]{Yes} (findMap);
            \draw[->, >=stealth] (findMap) --
            node[scale=1.3, xshift=1cm, yshift=-0.15cm]{$h: b \rightarrow r$} (confidentInMap);
            \draw[->, >=stealth] (confidentInMap) -- node[xshift=0cm, yshift=0.25cm]{Yes} (returnMap);

            \draw[->, >=stealth] (confidentInCatalog.west) -- ++(-1.4cm, 0cm) node[anchor=south, xshift=0.5cm]{No}
            |- (pickQueryStars.west);
            \draw[->, >=stealth] (confidentAfterFilter.west) -- ++(-1.4cm, 0cm) node[anchor=south, xshift=0.5cm]{No}
            |- (pickQueryStars.west);
            \draw[->, >=stealth] (confidentInMap.west) -- ++(-1.4cm, 0cm) node[anchor=south, xshift=0.5cm]{No}
            |- (pickQueryStars.west);
        \end{tikzpicture}
        \caption{
        Flowchart depicting the unified identification framework which all methods here follow.
        Given an image $I$, this process returns a bijection $h$ between some subset of the input $b$ and a subset of the
        catalog $r$.
        In the event all subsets are exhausted, the function $h: b \srightarrow \emptyset$ is returned (not depicted).
        } \label{figure:unifiedIdentificationFlowchart}
        }
    \end{figure}
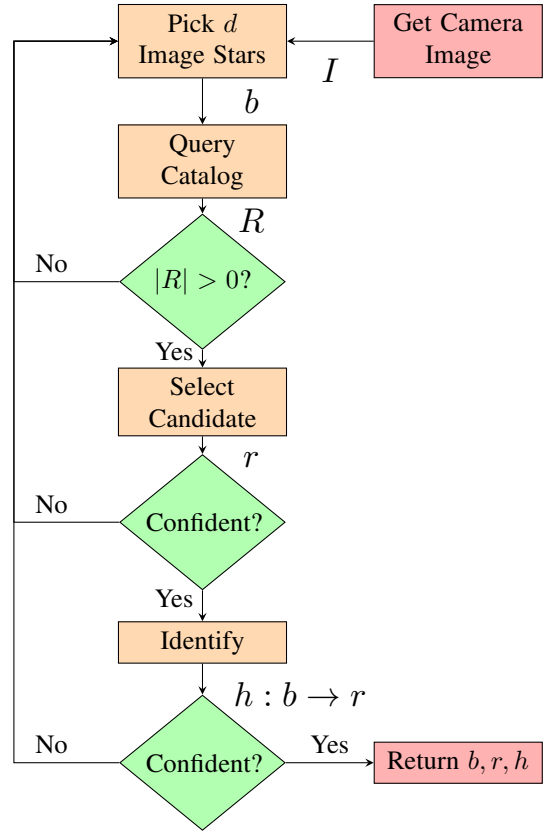

    \nsubparagraph{Database Access:}
    The naive approach to searching for matching features in a subgraph isomorphism approach is to perform a
    linear search across the entire catalog and search for matching subsets.
    Early star identification methods focused on reducing the size of the database to be queried, rather the query process
    itself.
    In 1996, Quine (according to Spratling) was the first to reduce the database search time from linear to log time
    using a binary search tree~\cite{quine:fastAutonomousStarAcquistion}.
    The following year Mortari's "Search-Less Algorithm" was introduced, which utilizes $k$-vectors to search the database
    independent of its size~\cite{mortari:kVectorApproach}.

    \nsubparagraph{Mapping:}
    To identify a star in an image is to pair it with some star in a catalog.
    Gottlieb's method used a voting approach to remove the ambiguity after identifying a single star
    pair~\cite{gottlieb:spacecraftAttitudeDetermination},
    which was later generalized by Kolomenkin to vote for every star in the image~\cite{kolomenkin:geometricVoting}.
    The direct match test was proposed by Needleman (according to Tappe), which determines the likelihood of a map based
    on how many stars from each frame align with the attitude formed by the map~\cite{needelman:stellarAttitudeAcquisition}.
    In an effort to avoid the mapping processes above, Anderson (according to Spratling) proposed the use of storing
    permutations of star subsets instead of combinations at the expense of storage~\cite{anderson:autonomousStarSensing}.
    The use of neural networks~\cite{lindsey:neuralNetworkMethods,alvelda:neuralNetworkStar} and genetic
    algorithms~\cite{paladugu:geneticAlgorithms} have also been proposed to optimize the mapping process.

    \newcommand{\algorithmautorefname}{Algorithm}
    \algnewcommand{\LineComment}[1]{\State \(\triangleright\) #1}
    \algrenewcommand\algorithmicindent{0.35cm}
    \newcommand{\bigO}{\mathcal{O}}
    \newcommand{\abs}[1]{\left| #1 \right|}
    \newcommand{\set}[1]{\left\{ #1 \right\}}
    \MakeRobust{\Call}

    \newfloat{algorithm}{t}{lop}
    \restylefloat{algorithm}

    \section{Star Identification Methods}\label{sec:starIdentificationMethods}
    Six different approaches to star identification are described in this section.
    The majority of the literature specifying identification methods do not include pseudocode, rather they specify
    descriptions of specific processes used by each method.
    Each algorithm is composed of these processes, structured to follow a general identification flow.

    \subsection{Unified Identification Framework}\label{subsec:unifiedIdentification}
    Each identification method is presented with information about the camera hardware, an image $I$ of size $n$
    containing all the stars in the image reference, as well as a catalog of known stars $K$.
    All stars in $I$ exist in the body frame $\iFrame$, and all stars in $K$ exist in the inertial frame $\kFrame$.
    The goal of each method is to find some bijection between a subset of the image stars $b$ and a subset
    of the catalog stars $r$.
    This function is denoted as $h$ with domain $b$ and codomain $r$.
    Identification of all stars in each image is not the focus.

    \begin{subequations}
        Every algorithm starts with some combination from all possible $d$ combinations of $n$ stars $C(n, d)$, where
        $d$ is the size of the image subset that specific identification method uses.
        $b$ is selected using one of these combinations.
        For an identification method that uses $d\seq2$ stars to determine the mapping in an image of $n\seq 4$ stars,
        the combinations of $I$ are:
        \begin{align}
            C(4, 2) \text{ of } I &= \set{\set{\vv{I_1}, \vv{I_2}}, \set{\vv{I_1}, \vv{I_3}}, \ldots,
            \set{\vv{I_3}, \vv{I_4}}} \\
            C(4, 2) \text{ of } I &= \set{ b_1, b_2, \ldots, b_6 }
        \end{align}
    \end{subequations}

    \begin{subequations}
        There exists a set $K^d$, composed of $d$ sized sets of all possible combinations (or permutations) of stars
        from the catalog $K$.
        Using certain features of the image star subset, the entire $K^d$ set is filtered to a set of catalog star
        candidates $R$.
        This is known as the catalog query step.
        Referencing the same $d\seq 2$ identification method as before, an image subset $b \seq \set{\vv{I_1},
        \vv{I_2}}$ may yield the candidates in~\autoref{eq:catalogCandidateExample}:
        \begin{align}
            \label{eq:catalogCandidateExample}
            R &= \set{ \set{\vv{K_{104}}, \vv{K_{899}}}, \set{\vv{K_{7622}}, \vv{K_{7771}}}, \ldots) } \\
            R &= \set{ r_1, r_2, \ldots }
        \end{align}
    \end{subequations}

    Through some filter process or restriction criteria for $R$ itself, a single set $r$ from the catalog star candidates is
    eventually selected.
    This may require going through multiple catalog candidate sets and repeating the catalog query step.
    This is known as the catalog candidate $r$ selection step.
    For a process with the $R$ restriction criterion of $|R| \seq 1$, the following sequence
    of events may occurring before finding a single $r$ set.
    \begin{align*}
        &t = 1, \text{ query with } b^{(1)}, \text{ get } R^{(1)} = \{\vv{r_{11}}, \vv{r_{12}}, \ldots\}. \\
        &t = 2, \ \abs{ R^{(1)} } \neq 1, \text{ criterion not met. } \\
        &t = 3, \text{ choose new image subset } b^{(2)}. \\
        &t = 4, \text{ query with } b^{(2)}, \text{ get } R^{(2)} = \{\vv{r_{21}}\}. \\
        &t = 5, \ \abs{ R^{(2)} } = 1, \text{ criterion met. } \\
        &t = 6, \text{ return } r, r \in R^{(2)} (\text{sole element in } R^{(2)} ).
    \end{align*}

    From here, a bijection $h : b \srightarrow r$ is determined that maps each star found in the image star subset to a
    single star in the catalog candidate set.
    If we are not confident in $h$ at this point, another image star subset is chosen and the process is repeated.
    If we are confident in $h$, then $b$, $r$, and $h$ are returned.
    This process is detailed in~\autoref{figure:unifiedIdentificationFlowchart}.
    In the event no map is determined, an error is raised and the function $h : b \srightarrow \emptyset$ is returned
    instead.


    \subsection{Angle Method (ANG)}\label{subsec:angleMethod}
    \newcommand{\invalidBijection}{\If{$\forall \ \vv{b^\star}, \ \vv{b^\star} \in b \land h\left(\vv{b^\star}\right)
    \neq \emptyset$}}
    \begin{algorithm}
        \caption{Angle Identification Method} \label{algorithm:angleIdentification}
        \begin{algorithmic}[1]
            \Function{FPO}{$P$, $I$, $A$}
            \State $I' \gets$ stars in $I$ rotated by $A$
            \State $\bar{P} \gets $ \{$p \in P \ | \ \exists \ i \ (i \in I' \land \theta (i, p) < 3\sigma_o)$\}
            \State \textbf{return} $\bar{P}$ \Comment Stars in $P$ that \textit{overlay} with $I'$.
            \EndFunction
            \\
            \Function{DMT}{$b, r, I$}
            \State $H \gets $ all possible bijections of $b$ and $r$
            \State $P \gets $ all stars in catalog near $r$, $M \gets \emptyset$
            \For {$h \in H$}
            \State $A \gets $ \Call{TRIAD}{$h, b, r$}, $M_h \gets $ \Call{FPO}{$P, I, A$}
            \EndFor
            \If{$\forall \ M_h \left( M_h \in M \land \abs{M_h} = \abs{b} \right)$}
            \State \textbf{return} $h : b \rightarrow \emptyset $ \Comment Not confident in result.
            \Else
            \State \textbf{return} $h \in H$ associated with largest set in $M$
            \EndIf
            \EndFunction
            \\
            \Function{Identify}{$I, K^2$}
            \For{$i \gets 1 \text{\textbf{ to }} n$} \Comment Iterate through $C(n, 2)$.
            \For{$j \gets i + 1 \text{\textbf{ to }} n - 1$}
            \State $b \gets \left(\vv{b_i}, \vv{b_j}\right)$, $R \gets \{ r \mid r \in K^2 \land P_\theta(r, b) \}$
            \If{$\lvert R \rvert = 1$}
            \State $h \gets $ \Call{DMT}{$b, R_1, I$} 
            \invalidBijection
            \State \textbf{return} $b, r, h$
            \EndIf
            \EndIf
            \EndFor
            \EndFor
            \EndFunction
        \end{algorithmic}
    \end{algorithm}

    The \textit{Angle} method is composed of a naive $b$ image subset decision, angular features of pairs first described by
    Gottlieb's Polygon Angular Matching method, and a direct-match test for identification.

    Given a set of stars from the image $I$, $d \seq 2$ stars are selected to obtain the $b$ set.
    The selection order is governed by lines (2) and (3) in~\autoref{algorithm:angleIdentification}.
    This fixes the star $\vv{b_1}$ in $b$ for $n$ image star subset selections, while constantly changing
    $\vv{b_2}$ for every new $b$ choice.
    An example sequence of pairs is depicted below for $n \seq 3$ stars.
    \begin{equation}
        C(3, 2) \text{ of } I = \left(\set{\vv{I_1}, \vv{I_2}}, \set{\vv{I_1},\vv{I_3}}, \set{\vv{I_2},\vv{I_1}},
        \ldots \right)
    \end{equation}

    The catalog query step searches the $K^2$ catalog for pairs such that the angular separations of the catalog pairs
    are close to the angular separation of the image star subset~\cite{bratt:analysisStarIdentification}.
    For the image star subset, the origin of the angular separation calculation $\theta(b)$ is the focal point of the lens
    itself.
    For a catalog star candidate set, the origin of this calculation $\theta(r)$ is the center of the Earth.
    To obtain $R$, the predicate $P_\theta(r, b)$ is used to filter the $K^2$ catalog:
    \begin{equation}\label{eq:angleRequirement}
    P_{\theta}(r, b) : \left\lvert \theta(r) - \theta(b)\right\rvert < 3 \sigma_\theta
    \end{equation}
    where $\sigma_{\theta}$ represents the deviation of the uncertainty between the $\theta$ computation with star
    sensor measurements and the same $\theta$ computation with stars defined in the catalog.
    Assuming the noise follows a Gaussian distribution, it follows that 99.7\% of all true pairs will be within this range
    ~\cite{coleAndCrassidis:sphericalTriangleMethod}.

    Once the catalog candidates are obtained, the $\abs{R} \seq 1$ criterion is imposed, repeating this process until only
    one candidate exists in $R$.
    This sole element $\vv{R_1}$ is then selected to be $r$.

    \begin{subequations}
        To determine the most likely bijection $h$, we follow Tappe's implementation of the method and perform a
        \textit{direct match test} (DMT)~\cite{tappe:starTrackerDevelopment,needelman:stellarAttitudeAcquisition}.
        Given an image star pair $b$ and a catalog star pair $r$ for, the following is proposed:
        \begin{equation}
            h_1 : \vv{b_1} \mapsto \vv{r_1}, \vv{b_2} \mapsto \vv{r_2}
        \end{equation}
        Wahba's problem is then solved using the TRIAD method to obtain a rotation $A_1$ between the image and catalog
        frames.
        This process is repeated for the other possible permutation to obtain a second rotation $A_2$:
        \begin{equation}
            h_2 : \vv{b_1} \mapsto \vv{r_2}, \vv{b_2} \mapsto \vv{r_1}
        \end{equation}
    \end{subequations}
    The most likely attitude is determined by the \Call{FPO}{} method, which returns how many stars from $I$ align with
    $K$ given rotation $A_1$ or $A_2$.
    The bijection with the most stars is then returned.
    If all bijections return sets of size $d \seq 2$, then we are not confident in any of our choices and
    return the function $h: \vv{b} \rightarrow \emptyset $.


    Accessing the catalog is the most expensive operation for all of the identification methods.
    Consequently, the running time of this algorithm $T_{angle}$ can be described in terms of the number of queries and
    the number of entries that exist in the $K^2$ catalog.
    There exist $2n^2$ catalog accesses at worst, requiring two catalog accesses (query step and \Call{DMT}{} calls) for
    each combination of pairs in $I$.
    The $\log (m_2)$ term describes the number of comparisons until $r$ sets are found and are able to be returned.
    Given a B+ tree indexed database with $\abs{K^2} \seq m_2$ elements, no more than $\bigO \left( \log(m_2) \right)$
    comparisons are required~\cite{patel:advanceTreeStructures}.
    \begin{equation}\label{eq:complexityAngle}
    T_{angle} = \bigO\left( n^2 \cdot \log(m_2) \right)
    \end{equation}


    \subsection{Interior Angle Method (INT)}\label{subsec:interiorAngleMethod}
    \begin{algorithm}
        \caption{Interior Angle Identification Method} \label{algorithm:interiorAngleIdentification}
        \begin{algorithmic}[1]
            \Function{Identify}{$I, \bar{K^3}$}
            \For{$c \gets 1 \text{\textbf{ to }} n$}  \Comment Iterate through all of $I$.
            \State $\theta_I \gets \{\theta(\vv{b_c}, \vv{b_i}) \mid \vv{b_i} \in I\}$ \Comment $\theta$
            (all stars, $\vv{b_c}$).
            \State $\vv{b_{c1}} \gets \vv{b_i}$ associated with smallest $\theta$ in $\theta_I$
            \State $\vv{b_{c2}} \gets \vv{b_i}$ associated with 2nd smallest $\theta$ in $\theta_I$
            \State $b \gets \left(\vv{b_c}, \vv{b_{c1}}, \vv{b_{c2}}\right)$, $R \gets \set{ r \mid r \in \bar{K^3} \land
            P_{\theta, \phi}(r, b) }$
            \If{$\lvert R \rvert = 1$}
            \State \textbf{return} $b, r, h: \vv{b_c} \mapsto \vv{r_c}, \vv{b_{c1}} \mapsto \vv{r_{c1}}, $
            \State \ \ \ \ \ \ \ \ \ \ \ \ \ \ \ \ \ \ \ \  $\vv{b_{c2}} \rightarrow \vv{r_{c2}}$
            \EndIf
            \EndFor
            \EndFunction
        \end{algorithmic}
    \end{algorithm}

    The \textit{Interior Angle} method is composed of Liebe's feature set (two interstar angles, an interior angle),
    Liebe's $b$ image subset decision, and a permutation store for identification.

    Given a set of stars from the image $I$, a central star $\vv{b_c}$ is selected.
    A new central star selection does not involve generating permutations like the Angle method, rather it involves
    iterating through $I$ in a sequential manner.
    The two closest stars in the image to the central star are selected next, denoted as $\vv{b_{c1}}$ and $\vv{b_{c2}}$
    ~\cite{liebe:starTrackersAttitudeDetermination}.

    The catalog query step searches the $\bar{K^3}$ catalog for trios such that the features of the catalog trios are
    close to the same features of the image subset~\cite{bratt:analysisStarIdentification}.
    Unlike the Angle method's $K^d$ set, $\bar{K^d}$ is defined to be all \textit{permutations} of size $d$ rather than
    combinations.
    These features are defined as the angular separation between the first closest star and the central star
    ($\theta\left(\vv{b_{c1}}, \vv{b_c}\right)$ vs. $ \theta\left(\vv{r_{c1}}, \vv{r_c}\right)$), the
    angular separation between the second closest star and the central star, ($\theta\left(\vv{b_{c2}},
    \vv{b_c}\right) $ vs. $\theta\left(\vv{r_{c2}}, \vv{r_c}\right)$),
    and the angular separation between the two closest stars with the central star as the origin instead of the Earth or
    focal point ($\phi(b) $ vs. $ \phi(r)$).
    To obtain $R$, the predicate $P_{\theta, \phi}(r, b)$ is used to filter the $\bar{K^3}$ catalog:

    \begin{equation}
        \begin{aligned}
            P_{\theta, \phi} (r, b): \left\lvert \theta(\vv{r_{c1}}, \vv{r_c}) - \theta(\vv{b_{c1}}, \vv{b_c})\right\rvert
            &< 3 \sigma_{\theta} \ \land \\ \left\lvert \theta(\vv{r_{c2}}, \vv{r_c}) - \theta(\vv{b_{c2}},
            \vv{b_c})\right\rvert &< 3 \sigma_{\theta} \ \land \\ \left\lvert \phi(r) - \phi(b)\right\rvert &< 3
            \sigma_\phi \ \land \\\theta(\vv{r_{c1}}, \vv{r_c}) &< \theta(\vv{r_{c2}}, \vv{r_c})
        \end{aligned}
    \end{equation}
    where $\sigma_{\theta}$ and $\sigma_{\phi}$ represent the deviation of the uncertainty between the $\theta$ and $\phi$
    computations with the star sensor measurements and the same $\theta$ and $\phi$ computations with stars defined in the
    catalog.

    After finding some $R$ that meets the same $R$ criterion as the Angle method, the bijection:
    \begin{equation}
        h: \vv{b_1} \mapsto \vv{r_1}, \vv{b_2} \mapsto \vv{r_2}, \vv{b_3} \mapsto \vv{r_3}
    \end{equation}
    is constructed and returned.
    RezaToloei's implementation imposes the last term in predicate $P_{\theta, \phi}(r, b)$ at query
    time~\cite{toloei:compositeIdentification}, borrowing from Anderson (according to Spratling) by searching all
    permutations instead of combinations to remove the need for a star mapping
    procedure~\cite{anderson:autonomousStarSensing}.
    Storing permutations does increase the storage required for the $\bar{K^3}$ catalog though, which begs the question,
    ``Does this extra space aid in accuracy or runtime?''.

    %

    The running time of this algorithm $T_{dot}$ is depicted below, again described in terms of the number of queries
    and $\bar{K^3}$ catalog entries:
    \begin{equation}\label{eq:dotComplexity}
    T_{dot} = O\left( n \cdot \log(\bar{m_3}) \right)
    \end{equation}
    where $\bar{m_3}$ is the size of the $\bar{K^3}$ catalog.


    \subsection{Spherical Triangle Method (SPH)}\label{subsec:sphericalTriangleMethod}
    \begin{algorithm}
        \caption{Triangle Method Identification} \label{algorithm:triangleIdentification}
        \begin{algorithmic}[1]
            \Function{PartialMatch}{$R, \bar{R}$}
            \ForAll {$\bar{r} \in \bar{R}$}
            \LineComment $\bar{r}$ and $r$ share two stars.
            \If {$\exists \ r \ | \ (r \in R \land |r \cap \bar{r}| = 2)$}
            \State $R_{new} \gets \bar{R} \cup \{\bar{r}\}$
            \EndIf
            \EndFor
            \State \textbf{return} $R_{new}$
            \EndFunction
            \\
            \Function{Pivot}{$\vv{b_i}, \vv{b_j}, \vv{b_k}, R$}
            \State $b \gets (\vv{b_j}, \vv{b_j}, \vv{b_k})$, $\bar{R} \gets \set{ \bar{r} \mid \bar{r} \in K^3
            \land P_{a, \tau}(\bar{r}, b) }$
            \State $R' \gets $ \Call{PartialMatch}{$R, \bar{R}$}
            \If{$\abs{R'} = 1 \lor \abs{R'} = 0 $}
            \State \textbf{return} $R'$ \Comment $R'$ is either $\emptyset$ or a single $r$.
            \Else
            \State $\vv{\beta} \gets \text{an unused star in this pivot}$
            \State \textbf{return} \Call{Pivot}{$\vv{b_i}, \vv{b_j}, \vv{\beta}, R'$}
            \EndIf
            \EndFunction
            \\
            \Function{Identify}{$I, K^3$}
            \For{$i \gets 1 \text{\textbf{ to }} n$}  \Comment Iterate through $C(n, 3)$.
            \For{$j \gets i + 1 \text{\textbf{ to }} n - 1$}
            \For{$k \gets j + 1\text{\textbf{ to }} n - 2$}
            \State $b \gets \left(\vv{b_i}, \vv{b_j}, \vv{b_k}\right)$
            \State $R \gets \set{ r \mid r \in K^3
            \land P{a, \tau}(r, b) }$
            \If{$|R| \neq 1$} \Comment Pivot if necessary.
            \State $R \gets $ \Call{Pivot}{$\vv{b_i}, \vv{b_j}, \vv{b_k}, R$}
            \EndIf
            \If{$R \neq \emptyset$} \Comment Verify the pivot's success.
            \State $h \gets $ \Call{DMT}{$b, R_1, I$}
            \invalidBijection
            \State \textbf{return} $b, R_1, h$
            \EndIf
            \EndIf
            \EndFor
            \EndFor
            \EndFor
            \EndFunction
        \end{algorithmic}
    \end{algorithm}

    The \textit{Spherical Triangle} method is composed of Cole and Crassidus's spherical area and moment features, a
    naive $b$ image subset decision, Cole and Crassidus's candidate selection process, and a direct-match test to create
    the image to catalog bijection.

    Given a set of stars from the image $I$, $d \seq 3$ stars are selected to obtain the $b$ set in the same
    straightforward manner as the Angle method.
    For $C(n, 3)$ combinations the star $\vv{b_1}$ is fixed in $b$ for $n^2$ image star subset selections,
    the star $\vv{b_2}$ is fixed for $n$ selections, and the last star $\vv{b_3}$ is constantly changed for
    every new $b$ choice.

    The catalog query step searches the $K^3$ catalog for trios such that the spherical area and moment of the catalog trios
    are close to the spherical area and moment of the image star subset~\cite{coleAndCrassidis:sphericalTriangleMethod}.
    For the image star subset, the spherical area and moment are represented as $a(b)$ and $\tau(b)$ respectively.
    For the catalog star candidate set, these same features are represented as $a(r)$ and $\tau(r)$.
    To obtain $R$, the predicate $P_{a, \tau}(r, b)$ is used to filter the $K^3$ catalog:
    \begin{equation}
        \begin{aligned}
            P_{a, \tau}(r, b) : \abs{ a(r) - a(b)} &< 3\sigma_{a}
            \ \land \\ \abs{\tau(r) - \tau(b)} &< 3\sigma_{\tau}
        \end{aligned}
    \end{equation}
    where $\sigma_{a}$ and $\sigma_{\tau}$ represent the deviation of the uncertainty between the $a$ and $\tau$
    computations with the star sensor measurements and the same $a$ and $\tau$ computations with stars defined in the
    catalog.

    Unlike the previous two methods, the $R$ criterion of $|R| \seq 1$ not being met does not lead to an immediate new
    selection of $b$.
    Instead, the candidate set itself is reduced by \textit{pivoting} until the criterion is met or pivots can no longer
    be performed.
    The procedure starts by querying the catalog again for a second set of catalog candidate sets $\bar{R}$ with a
    different image star subset $\bar{b} \seq \left(\vv{b_i}, \vv{b_j}, \vv{\beta}\right)$.
    In $\bar{b}$, the first two stars are held constant while the third star is swapped with another in $I$ that was not
    already used in this specific pivot.
    All star trios in the initial search that do not match a trio in the second search by \textit{two stars} (a partial
    match) are removed from the initial search candidate star set~\cite{coleAndCrassidis:sphericalTriangleMethod}.
    A pivot uses at most $n - 3$ additional catalog accesses, but prevents wasting a catalog candidate set that may contain
    the correct $r$ set for the given $b$.

    \begin{subequations}
        The \Call{DMT}{} process is used to complete the star identification process here.
        Given an image star trio and a catalog star trio, a bijection is proposed:
        \begin{equation}
            h_1 : \vv{b_1} \mapsto \vv{r_1}, \vv{b_2} \mapsto \vv{r_2}, \vv{b_3} \mapsto \vv{r_3}
        \end{equation} \label{eq:trianglePossibleMaps}
        The TRIAD method only uses two vector observations from each frame, meaning that the $\vv{b_3} \rightarrow \vv{r_3}$
        pairing is disregarded as the first rotation $A_1$ is computed.
        This process is repeated for all 5 other possible bijections to get $A_2, A_3, \dots, A_6$.
        \begin{align}
            h_2 &: \vv{b_1} \mapsto \vv{r_1}, \vv{b_2} \mapsto \vv{r_3}, \vv{b_3} \mapsto \vv{r_2} \\
            h_3 &: \vv{b_1} \mapsto \vv{r_2}, \vv{b_2} \mapsto \vv{r_1}, \vv{b_3} \mapsto \vv{r_3} \\
            h_4 &: \vv{b_1} \mapsto \vv{r_2}, \vv{b_2} \mapsto \vv{r_3}, \vv{b_3} \mapsto \vv{r_1} \\
            h_5 &: \vv{b_1} \mapsto \vv{r_3}, \vv{b_2} \mapsto \vv{r_1}, \vv{b_3} \mapsto \vv{r_2} \\
            h_6 &: \vv{b_1} \mapsto \vv{r_3}, \vv{b_2} \mapsto \vv{r_2}, \vv{b_3} \mapsto \vv{r_1}
        \end{align}
    \end{subequations}
    For all six attitudes, the bijection yielding the most aligned stars is returned.

    The running time of this algorithm $T_{sphere}$ is depicted below in terms of the number of queries and the number of
    entries in the $K^3$ catalog.
    At most, this requires $2n^4$ catalog access: $n^3$ for each combination of trios in $I$, $n - 3$ potential catalog
    accesses incurred for each pivot, and an additional $n^4$ queries with each \Call{DMT}{} call.
    \begin{equation}\label{eq:sphereComplexity}
    T_{sphere} = \bigO\left( n^4 \cdot \log (m_3) \right)
    \end{equation}
    where $m_3$ represents the size of the $K^3$ catalog.


    \subsection{Planar Triangle Method (PLN)}\label{subsec:coleAndCrassidus'sPlanarTriangleMethod}
    The \textit{Planar Triangle} method is identical to their Spherical Triangle method, with the exception that each image
    trio is represented as a planar triangle instead of a spherical one.
    This results in the computation of a planar area and moment as opposed to a spherical area and moment.

    \subsection{Pyramid Method (PYR)}\label{subsec:pyramidMethod}
    \begin{algorithm}
        \caption{Pyramid Identification Method} \label{algorithm:pyramidIdentification}
        \begin{algorithmic}[1]
            \Function{FC}{$T_1, T_2$}
            \LineComment Flatten $T_1, T_2$ from set of sets to a set.
            \State $\bar{T_1} \gets \emptyset, \bar{T_2} \gets \emptyset$
            \ForAll{$i \in \set{1, 2}$}
            \ForAll{$\vv{t} \in T_i$}
            \State $\bar{T_i} \gets \bar{T_i} \cup \set{ \vv{t_1}, \vv{t_2} }$
            \EndFor
            \EndFor
            \State \textbf{return} $\bar{T_1} \cap \bar{T_2}$
            \EndFunction
            \\
            \Function{FindT}{$\vv{b^1}, \vv{b^2}, \vv{b^3}, K^2$}
            \LineComment $\abs{\vv{b^1}} = \abs{\vv{b^2}} = \abs{\vv{b^3}} = 1$, search with each pair.
            \State $T_1 \gets \set{ r \mid r \in K^2 \land P_\theta \left(r, b^1\right) }$
            \State $T_2 \gets \set{ r \mid r \in K^2 \land P_\theta \left(r, b^2\right) }$
            \State $T_3 \gets \set{ r \mid r \in K^2 \land P_\theta \left(r, b^3\right) }$
            \State \textbf{return}{$\left( T_1, T_2, T_3 \right)$}
            \EndFunction
            \\
            \Function {Query}{$b, K^2$}
            \State $(T_{ij}, T_{ik}, T_{jk}) \gets$ \Call{FindT}{$\set{ \vv{b_i}, \vv{b_j} }, \set{ \vv{b_i}, \vv{b_k} }, $
            \State \ \ \ \ \ \ \ \ \ \ \ \ \ \ \ \ \ \ \ \ \ \ \ \ \ \ \ \ \ \ \ \
        $\set{ \vv{b_j}, \vv{b_k} }, K^2$}, $R \gets \emptyset$
            \ForAll{$t_i \in$ \Call{FC}{$T_{ij}, T_{ik}$}}
            \ForAll{$t_j \in$ \Call{FC}{$T_{ij}, T_{jk}$}}
            \ForAll{$t_k \in$ \Call{FC}{$T_{ik}, T_{jk}$}}
            \State $R \gets R \cup \set{ \left(\vv{t_i}, \vv{t_j}, \vv{t_k}\right) }$
            \EndFor
            \EndFor
            \EndFor
            \State \textbf{return} $R$ \Comment Return all permutations from $T$ sets.
            \EndFunction
            \\
            \Function{Identify}{$I, K^2$}
            \LineComment Iterate through $C(n, 3)$ while avoiding false stars.
            \For{$dj \gets 1 \text{\textbf{ to }} n - 2$}
            \For{$dk \gets 1 \text{\textbf{ to }} n - 1 - dj$}
            \For{$i \gets 1 \text{\textbf{ to }} n - dj - dk$}
            \State $j \gets i + dj$, $k \gets j + dk$
            \State $b \gets \left(\vv{b_i}, \vv{b_j}, \vv{b_k}\right)$, $R \gets $\Call{Query}{$b, K^2$}
            \If{$\abs{R} = 1$}
            \LineComment Verification step below.
            \State $\vv{\beta} \gets $ single star in $I$ where $\vv{\beta} \notin b$
            \State $(T_{ij}, T_{ik}, T_{jk}) \gets$ \Call{FindT}{$\set{ \vv{b_i}, \vv{\beta} }, \set{ \vv{b_j},
            \vv{\beta} },$
            \State \ \ \ \ \ \ \ \ \ \ \ \ \ \ \ \ \ \ \ \ \ \ \ \  \ \ \ \ \ \ \ \ $\set{ \vv{b_k}, \vv{\beta} }, K^2$}
            \State $T_\beta \gets $ \Call{FC}{$T_{i\beta}, T_{j\beta}$} $\cap$ \Call{FC}{$T_{j\beta}, T_{k\beta}$}
            \If{ $\abs{T_\beta} = 1$}
            \State \textbf{return} $b, r, h: \vv{b_1} \mapsto \vv{r_1}, \vv{b_2} \mapsto \vv{r_2},$
            \State \ \ \ \ \ \ \ \ \ \ \ \ \ \ \ \ \ \ \ \ $\vv{b_3} \mapsto \vv{r_3}$
            \EndIf
            \EndIf
            \EndFor
            \EndFor
            \EndFor
            \EndFunction
        \end{algorithmic}
    \end{algorithm}

    The \textit{Pyramid} method is composed of Mortari's $b$ image subset decision, a custom voting based identification
    process for star trios, and a voting based verification step.

    Given a set of stars from the image $I$, $d \seq 3$ stars are selected to obtain the $b$ set.
    The selection order is governed by lines (27), (28), and (29) in~\autoref{algorithm:pyramidIdentification}.
    As opposed to the selection order of the Angle and triangle methods, the $\vv{b_1}$ star in $b$ is no longer fixed
    for $n$ or $n^2$ image star subset selections.
    This is meant to avoid the persistence of misleading stars for more than a few combinations
    ~\cite{mortari:pyramidIdentification}.
    An example sequence of trios is depicted below for $n \seq 5$ stars.
    \begin{equation}
        \begin{aligned}
            C(5, 3) \text{ of } I = ( &\set{\vv{I_1}, \vv{I_2}, \vv{I_3}}, \set{\vv{I_2}, \vv{I_3}, \vv{I_4}}, \\
            &\set{\vv{I_3}, \vv{I_4}, \vv{I_5}}, \set{\vv{I_1}, \vv{I_2}, \vv{I_4} }\ldots )
        \end{aligned}
    \end{equation}

    The approach developed here was inspired by the two star voting algorithm, which accumlates "votes" for some star by
    determining the angle between the same star and two other stars~\cite{tichy:preliminaryTestsCommericalImagers}.
    We start by querying for pairs from the $K^2$ catalog such that the angular separations of the catalog pairs are
    close to the angular separation of the image pair $\set{\vv{b_i}, \vv{b_j}}$.
    This is repeated for the other two permutations $\set{\vv{b_i}, \vv{b_k}}$ and $\set{ b_j, \vv{b_k} }$ to obtain
    the sets $T_{ij}, T_{ik}$ and $T_{jk}$ respectively.
    These sets are then flattened from sets of pairs to just a single set of stars (\Call{FC}{}
    in~\autoref{algorithm:pyramidIdentification}) and the difference of two flattened sets identify candidates for that
    star.
    For the sets of pairs $T_{ij}, T_{jk}$ found by querying with $P_\theta$ and $\set{\vv{b_i}, \vv{b_j}},
    \set{\vv{b_j}, \vv{b_k}}$, the common star between each $b$ set is $\vv{b_j}$.
    An example of finding catalog candidates for $b_j$ with this method is given below:
    \begin{equation}
        \begin{aligned}
            T_{ij} \gets& \set{ \set{\vv{K_{1123}}, \vv{K_{9001}}}, \set{\vv{K_{8234}}, \vv{K_{33}}} } \\
            T_{jk} \gets& \set{ \set{\vv{K_{612}}, \vv{K_{1123}}}, \set{\vv{K_{33}}, \vv{K_{345}}} } \\
            T_j =& \  \Call{FC}{T_{ij}, T_{jk}} = \set{\vv{K_{1123}}, \vv{K_{33}}}
        \end{aligned}
    \end{equation}
    $R$ is found by repeating the process above for $T_i$ and $T_k$, and generating all possible sequences.
    This is depicted in the \Call{Query}{} function in~\autoref{algorithm:pyramidIdentification}.

    After finding some $R$ where $\abs{T_i} \seq \abs{T_j} \seq \abs{T_k} \seq 1$ (same criterion as Angle and
    Interior Angle), a verification step is performed.
    A different star from the image $\vv{\beta}$ is selected and the query step is performed for each distinct trio
    combination of $b$ and $\vv{\beta}$.
    If $\abs{T_\beta} \neq 1$, then verification step has failed and another image subset is selected.
    Otherwise, the bijection $h : \vv{b_1} \mapsto \vv{r_1}, \vv{b_2} \mapsto \vv{r_2}, \vv{b_3} \mapsto \vv{r_3}$ is
    returned.
    Like the Interior Angle method, a star mapping procedure is not required to determine $h$.
    Instead, each individual star is identified at query time.\looseness=-1

    The running time of this algorithm $T_{pyramid}$ is depicted below in terms of the number of queries and the number
    of entries in the $K^2$ catalog.
    At most, this requires $6n^3$ catalog accesses: $3n^3$ accesses for each query step with an additional $3n^3$ accesses
    for each verification step.
    \begin{equation}
        T_{pyramid} = \bigO \left( n^3 \cdot \log( m_2 ) \right)
    \end{equation}
    where $m_2$ is the size of the $K^2$ catalog.


    \begin{algorithm}
        \caption{Composite Pyramid Identification Method}\label{algorithm:compositePyramid}
        \begin{algorithmic}[1]
            \Function{Identify}{$I, K^3$}
            \LineComment Iterate through $C(n, 3)$ while avoiding false stars.
            \For{$dj \gets 1 \text{\textbf{ to }} n - 2$}
            \For{$dk \gets 1 \text{\textbf{ to }} n - 1 - dj$}
            \For{$i \gets 1 \text{\textbf{ to }} n - dj - dk$}
            \State $j \gets i + dj$, $k \gets j + dk$
            \State $b \gets (\vv{b_i}, \vv{b_j}, \vv{b_k})$
            \State $R \gets \set{ r \mid r \in K^3 \land P{a, \tau}(r, b) }$
            \If{$\lvert R \rvert = 1$}
            \LineComment Verification step below.
            \State $\beta \gets $ single star in $I$ where $\beta \notin b$
            \State $T_{12\beta} \gets \set{ r \mid r \in K^3 \land P_{a, \tau}(r, \set{\vv{b_1}, \vv{b_2}, \vv{\beta}})}$
            \State $T_{13\beta} \gets \set{ r \mid r \in K^3 \land P_{a, \tau}(r, \set{\vv{b_1}, \vv{b_3}, \vv{\beta}})}$
            \State $T_{23\beta} \gets \set{ r \mid r \in K^3 \land P_{a, \tau} (r, \set{\vv{b_2}, \vv{b_3}, \vv{\beta}})}$
            \State $T_\beta \gets $ \Call{FC}{$T_{12\beta}, T_{13\beta}$} $\cap$ \Call{FC}{$T_{j13\beta}, T_{23\beta}$}
            \If{$\abs{T_\beta} = 1$}
            \State $h \gets$ \Call{DMT}{$b, R_1, I$}
            \invalidBijection
            \State \textbf{return} $h$
            \EndIf
            \EndIf
            \EndIf
            \EndFor
            \EndFor
            \EndFor
            \EndFunction
        \end{algorithmic}
    \end{algorithm}

    \subsection{Composite Pyramid Method (COM)}\label{subsec:compositePyramidMethod}
    The \textit{Composite Pyramid} method is composed of Mortari's $b$ image subset decision, Cole and Crassidus's
    spherical area and moment features, and a voting based verification step.

    Given a set of stars from the image $I$, $d \seq 3$ stars are selected in same manner as the Pyramid method to obtain
    the $b$ set.
    From here, the process to obtain the $R$ set is the same as the triangle methods: use $P_{a, \tau}(r, b)$ and $b$ to
    select all candidates from $K^3$.
    If the current $R$ set meets the same $\abs{R} \seq 1$ criterion, then a similar verification step to the Pyramid method
    is performed with the Planar Triangle features.
    Once this test has passed, the \Call{DMT}{} method is used to construct the bijection $h$ to potentially return.
    The Pyramid method did not need this call as an implicit bijection was formed through its query process.

    The running time of this algorithm $T_{composite}$ is depicted below in terms of number of queries and the number of
    items in the $K^3$ catalog.
    At most, this requires $5n^3$ catalog accesses: $n^3$ for each query step, an additional $3n^3$ accesses for each
    verification step, and an additional $n^3$ accesses for each \Call{DMT}{} call.
    \begin{equation}
        T_{composite} = \bigO (n^3 \cdot \log(m_3))
    \end{equation}
    where $m_3$ represents the number of entries in the $K^3$ catalog.



    \newcommand{\AVG}{\mathit{AVG}}

    \section{Empirical Evaluation}\label{sec:empiricalEvaluation}
    In this section all six identification methods are analyzed in terms of their process to obtain the catalog candidate
    set $R$ (query step), their catalog set $r$ selection process, and their bijection $h$ production process
    (identification) under varying amounts of false stars and Gaussian noise.
    The main areas of interest here are the accuracy of each step, and the time to produce a result.


    \subsection{Experimental Setup}\label{subsec:experimentalSetup}
    \nsubparagraph{Star Catalog:}
    The star catalog used for $K$ is the Hipparcos Input Catalogue~\cite{perryman:hipparcosCatalogue}.
    Entries that do not have a point $\left( \alpha, \delta \right)$ associated with it were not recorded, giving
    $117{,}956$ total stars.
    Out of this entire set, only $4{,}560$ are visible from Earth with the naked eye (apparent magnitude $m$ less than 6.0).
    An additional constraint for each catalog $K^2, K^3, \bar{K^3}$ that all stars in each pair or trio be within 20
    degrees of each other was placed to shorten each algorithm's query step running time.
    A field-of-view between 10 to 20 degrees is common for most astronomy based CCD
    cameras~\cite{mortari:pyramidIdentification}.
    All sets $K^2, K^3, \bar{K^3}$ construct combinations and permutations using the $4{,}560$ elements and this field
    of view constraint.
    To construct the point $[ x \ y \ z ]$ for $K$,~\autoref{eq:sphereToCartesian} was used with
    each recorded $\left(\alpha, \delta \right)$ and $r \seq 1$, then normalized.

    \nsubparagraph{Benchmark Data Generation:}
    Before a raw image can be used in any of the star identification algorithms presented above, it must go through
    three major processes: blob detection, centroid determination, and a 2D $\rightarrow$ 3D transformation process.
    If a blob is not wholly detected, the centroid is not determined correctly, or the transformation process
    is not precise enough, error will exist as input to the algorithm prior to starting.
    Given that our goal is to only characterize each star identification algorithm itself, the solution implemented here
    involves generating artificial images in some quasi 3D space.

    Prior to generating the benchmark data, three items are specified: a field of view $\psi$, a true attitude
    $A^{\nicefrac{\iFrame}{\kFrame}}$, and a 3D vector $\vv{r_f}$ in the catalog frame $\kFrame$ that determines
    the center of the image.
    The next step is to find all nearby stars to the $\vv{r_f}$ in the catalog.
    This is denoted as $J$:
    \begin{equation}
        J = \set{ j \mid j \in K \land \theta\left( j, \vv{r_f} \right) < \frac{\psi}{2} }
    \end{equation}
    To get the $I$ set, each star in $J$ is then rotated by the true attitude $A^{\nicefrac{\iFrame}{\kFrame}}$:
    \begin{equation}
        I = \set{ A^{\nicefrac{\iFrame}{\kFrame}} \cdot j \mid j \in J }
    \end{equation}
    The set $I$, the field of view, and the rotated image center
    $\vv{b_f} \seq A^{\nicefrac{\iFrame}{\kFrame}} \cdot \vv{r_f}$ are then presented to each star identification algorithm.

    The first type of noise exists as variance between the relative positions of stars represented in the catalog and those
    represented in the image.
    This may come from misidentifying the centroids in the image or out-of-date catalogs.
    To introduce Gaussian noise to an image, we spherically linearly interpolate each star toward some random 3D vector on
    the unit sphere (\textit{SLERP}) and distribute the magnitude of the movement normally.
    To describe our noise independent of this random vector, we divide a normal random variable by the current angular
    separation between both stars.
    Given a star $b_i\!\in\!I$, Gaussian noise is applied to obtain the distributed vector $b'_i$~\cite{kremer:slerp}:
    \begin{equation}
        b'_i = \frac{\sin (1 - K)\Omega}{\sin \Omega}b_i + \frac{\sin \left( K \Omega \right)}{\sin \Omega}b^\star_i
    \end{equation}
    \begin{subequations}
        where $b^\star_i$ represents some random vector with uniformly distributed elements, $\Omega$ describes the
        angle subtended by the arc, and $K$ describes the magnitude of the interpolation.
        Below, $\rho$ represents the standard deviation of noise.
        \begin{align}
            b^\star_i &= \left[ \sim U(-1, 1), \sim U(-1, 1), \sim U(-1, 1) \right] \\
            \Omega &= \arccos \left ( b^{\star}_i \cdot b_i \right) \\
            K &= \left(\sim N\left(0, \rho^2\right)\right) \cdot \left(\theta\left( b^{\star}_i, b_i \right)
            \right)^{-1}
        \end{align}
        The additional constraint that the resulting star exist near the image center is also applied:
        $\theta\left( b'_i, \vv{r_f} \right)\!<\!\nicefrac{\psi}{2}$.
        If this is not met, then the process is repeated for this star.
    \end{subequations}

    The second type of noise exists as falsely identified sources of light, or spikes in the image.
    This involves generating $b^\star_i$ in the same manner that was done for the Gaussian noise process, and normalizing
    this.
    If the constraint that $b^\star_i$ be near the image center is not met, this process is repeated until such a star is
    found.
    This is repeated for a set number of spikes.

    \nsubparagraph{Hardware:}
    All trials were performed on an Intel i7-7700 CPU, 3.60GHz with 8 GB RAM\@.
    Each algorithm was implemented in C++14, and compiled without optimization (at \texttt{-O0}).
    The exact implementation is available at the following link:
    \url{https://github.com/glennga/hoku}.

    \subsection{Catalog Query Step}\label{subsec:catalogQueryStep}
    \nsubparagraph{Determining Query $\sigma$:}
    In all predicates used to query the catalog, an assumption must be made about the difference between the catalog
    measurements and the image measurements.
    If this deviation assumption $\sigma$ is too large, false positives will exist in $R$ after querying and may slow
    down identification.
    On the other hand, $\abs{R} \seq 0$ if the deviation assumption is too small.
    The heuristic used to determine each query $\sigma$ was to exhaust every permutation of deviations in the set below for
    30 query steps each.
    Work toward more accurately estimating star identification parameters has been performed by
    Balodis~\cite{balodis:parametersAutomated}:
    \begin{equation}
        \sigma_{gd} \in \set{ 10^{-16}, 10^{-15}, \ldots, 10^1 }
    \end{equation}
    The Interior Angle and triangular feature based methods of $\abs{\omega} \seq 2$ have $18^2$ distinct parameter sets
    with 30 runs attached to each set.
    The Angle and Pyramid method of $\abs{\omega} \seq 1$ has $18$ distinct parameter sets with 30 runs attached to each set.
    The parameter sets with the largest $\sigma$ choices but most number of instances where $\abs{R} \seq 1$ were selected.

    The results for each method are displayed below, and were used for the following experiments.
    \begin{alignat*}{3}
        \text{ANG / PYR}&: \sigma_\theta &&= 10^{-4} &&{} \\
        \text{INT}&: \sigma_\theta &&= 10^{-2}, \sigma_\phi &&= 10^{-2} \\
        \text{SPH / PLN / COM}&: \sigma_a &&= 10^{-9}, \sigma_\tau &&= 10^{-9}
    \end{alignat*}

    \begin{table}
        \centering {

        \begin{tabular}{m{0.22\columnwidth}|m{0.2\columnwidth}|m{0.2\columnwidth}|m{0.2\columnwidth}}
            \toprule
            \textit{Method} & $f_{r_b \in R}$ & $S$ & $t_{\AVG} \ (\si{ms})$  \\ \hline
            ANG & \num{1.0} & \num{32} & \num{138.00} \\ \hline
            INT & \num{1.0} & \num{1440} & \num{171.80} \\ \hline
            PLN / COM & \num{1.0} & \num{1994} & \num{139.05} \\ \hline
            SPH & \num{1.0} & \num{1984} & \num{139.60} \\ \hline
            PYR & \num{0.99} & \num{1501} & \num{149.69} \\ \bottomrule
        \end{tabular}
        \caption{
        Depicts all data associated with testing the query step: the frequency of correct catalog sets ($r_b$,
        such that the correct bijection can be formed with $b$) existing in $R$ after querying, the number of trials
        where the resulting $R$ meets the $\abs{R} \seq 1$ criterion ($S$),
        and the average query running time ($t_{\AVG}$) given images with no noise.
        There exist $2{,}000$ runs for each identification method.
        } \label{tab:queryExperimentResults}
        }
    \end{table}

    \subsubsection{Which method has the fastest catalog query step?}
    In~\autoref{sec:starIdentificationMethods}, we describe each method's running time in terms of the number of catalog
    accesses $n$ and the size of the $K^d$ catalog.
    The $K^2$ catalog, used by the Angle and Pyramid methods, is of size $m_2 \seq 353{,}700$ elements with the apparent
    magnitude and field-of-view constraints.
    The $K^3$ catalog, used by the Spherical Triangle, Planar Triangle, and Composite Pyramid methods is of size
    $m_3 \seq 12{,}520{,}359$ elements.
    The $\bar{K^3}$ catalog, used by the Interior Angle method is of size $\bar{m_3} \seq 37{,}561{,}083$ elements.
    Given the size of each catalog, we expect that the Angle method will have the fastest query step and the Interior Angle
    will have the slowest query step.

    \begin{figure}
        \begin{align*}
            \texttt{SELECT } &r \\
            \texttt{FROM } &K^d \\
            \texttt{WHERE } &g_1(r) < g_1(b) + 3\sigma_{g1} \texttt{ AND } \\
            &g_1(r) > g_1(b) - 3\sigma_{g1} \texttt{ AND } \\
            &g_2(r) < g_2(b) + 3\sigma_{g2} \texttt{ AND } \\
            &g_2(r) > g_2(b) - 3\sigma_{g2} \texttt{ AND } \\
            &\vdots \\
            &g_d(r) < g_d(b) + 3\sigma_{gd} \texttt{ AND } \\
            &g_d(r) > g_d(b) - 3\sigma_{gd}
        \end{align*}
        \caption{
        Depicts a generalized SQL query used for the Angle, Spherical Triangle, Planar Triangle, and Composite Pyramid
        methods.
        Here, $d$ represents the number of stars used in the query, $g$ represents the function used to obtain a feature,
        and $\sigma$ refers to the deviation of noise.
        }\label{fig:sqlQuery}
    \end{figure}

    In~\autoref{tab:queryExperimentResults}, the average running time to obtain an $R$ set is displayed for each
    identification method given an image for $2{,}000$ runs.
    The slowest method on average is the Interior Angle method, with its $t_{\AVG} = 30.64 \si{ms}$ longer than the average
    $t_{\AVG}$ for all other methods ($141.16 \pm 4.30 \si{ms}$).
    More time is being spent searching for the appropriate elements.

    We note that the two fastest methods appear to be Angle method and the Planar Triangle method, but their $t_\AVG$ only
    vary by 1.05ms.
    Given the null hypothesis that the difference between the Planar Triangle method's query step running time and the
    Angle method's query step running time is not significant, $z \seq 8.75, p\!<\!0.0001$ is found with a two-tailed two
    sample $Z$ test.
    The Angle method has the fastest query step due its small catalog size.

    \subsubsection{Which method meets the $\abs{R} \seq 1$ criterion the most often?}
    The $\abs{R} = 1$ criterion is required for all identification methods at some point (after pivoting for the triangle
    methods), and meeting this criteria as often as possible prevents additional catalog accesses from occurring.

    In~\autoref{tab:queryExperimentResults}, the lowest number of instances where the criterion is met $S$ lies with the
    Angle method.
    Out of $2{,}000$ query steps, the Angle method will have had to perform an additional query step at least
    $1{,}968$ more times.
    The Pyramid method only has 499 of these additional query instances, which is a factor of 3.94 less.
    The most likely reason for this lies with the selection of the $\sigma_\theta$ parameter, and the fact that only one
    feature is used to query $K^2$.
    this comes at the cost of being less flexible with Gaussian noise.
    The methods using $K^3$ and $\bar{K^3}$ have the advantage of being able to create utilize more features of the $b$ set
    and distinguish it better, compared to only using $\theta(b, r)$ as the sole feature.

    It appears that the all methods using triangular features (Planar Triangle, Composite Pyramid, Spherical Triangle)
    meet the criterion the most often (average of $1{,}989.7 \pm 4.2$ runs).
    Again, a larger $\sigma_a$ or $\sigma_\tau$ query parameter may lead to a larger $\abs{R}$.
    The next method with the most $\abs{R} \seq 1$ runs that does not use triangular features is the Pyramid method,
    which has a factor of 0.75 less runs.
    Methods with triangular features are more likely on average to have more instances where the $R$ criterion is met
    when compared to methods with angular features.

    \enlargethispage{-\baselineskip} \enlargethispage{-\baselineskip}
    \subsubsection{How effective is the Pyramid method query?}
    In the Angle, Spherical Triangle, Planar Triangle, and Composite Pyramid methods, catalog queries can be
    generalized to the query in~\autoref{fig:sqlQuery}.
    The Interior Angle method requires the $\theta(r_{c1}, r_{c})\!<\!\theta(r_{c2}, r_c)$ constraint before performing the
    query above.
    Compared to the rest of the methods presented here, the Pyramid method has the most involved query that involves
    processing outside of SQL\@.
    Three of the queries above must be performed to obtain the $T$ sets, and the common stars must be
    found among each $R$ set to create a singular candidate set for trios.


    The additional complexity of the Pyramid method increases the frequency of false negatives after querying.
    In~\autoref{tab:queryExperimentResults}, the frequency of the correct $r$ existing in $R$ for some $b$ is displayed
    for each identification method.
    The Pyramid method is shown to have a 0.01\% difference from the 100\% accuracy of each other method.
    Given the null hypothesis that this difference is not significant, $z \seq 4.49, p\!<\!0.0001$ is obtained with a
    one tailed two sample $Z$ test.
    We find that the Pyramid method's query step is less accurate than other identification methods.
    Although small, this error will propagate to the next steps and will result in more catalog accesses and/or a lower
    average accuracy.

    \subsection{Candidate Selection Step}\label{subsec:candidateSelectionStep}
    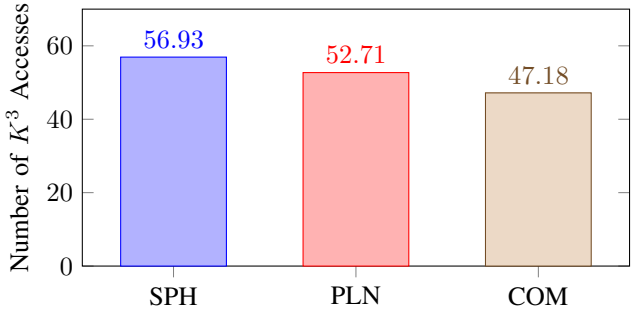
\begin{figure}
        \centering{
        \begin{tikzpicture}
            \begin{axis}[
            ybar,
            width=\linewidth, height=5cm,
            ylabel={Number of $K^3$ Accesses}, ylabel near ticks, ymin=0, ymax=70,
            xticklabels={SPH, PLN, COM},
            xtick={1, 2, 3}, xmin=0.5, xmax=3.5, xtick pos=left,
            nodes near coords, nodes near coords align={vertical},
            every axis plot/.append style={
            ybar,
            bar width=40,
            bar shift=0pt,
            fill
            }
            ]
                \addplot coordinates {(1, 56.93)}; 
                \addplot coordinates {(2, 52.71)}; 
                \addplot coordinates {(3, 47.18)};
            \end{axis}
        \end{tikzpicture}


        \caption{
        Depicts the average number of catalog accesses required to obtain a $r$ set for methods with triangular
        features given $\rho \seq \ang{0.0001}$ of Gaussian noise.
        To characterize the pivoting method itself, we only display instances where $\abs{R}\!\neq\!1$ with the first $b$
        selection.
        The Spherical Triangle method has $1{,}952 / 2{,}000$ runs matching the criteria before, the Planar Triangle
        method has $1{,}946$ runs, and the Composite Pyramid method has $1{,}957$ runs.
        }\label{fig:rPivot}
        }
    \end{figure}

    \subsubsection{How expensive is the pivoting process?}
    As seen previously, identification methods with triangular features have the most number of instances where
    $\abs{R} \seq 1$ given an image with no noise.
    ~\autoref{fig:rPivot} displays the average number of catalog accesses for these same methods where the first
    $b$ selection does not meet the $R$ criterion given an image with Gaussian noise.
    We note that the average number of catalog accesses is higher in methods that use the pivoting processes,
    as opposed to those that do not.
    Given the null hypothesis that the difference between the Planar Triangle method's number of catalog accesses and
    the Composite Pyramid method's number of catalog accesses is not significant, $z \seq 3.3, p\!<\!0.0001$ is
    obtained with a two-tailed two sample $Z$ test.
    With the data collected here, we find that the pivoting process results in more catalog accesses on average.
    This increased number of catalog accesses results in a $6.70\si{ms}$ difference on average between the two.

    The pivoting process was only tested with the methods most frequently meeting the $R$ criterion.
    An area of interest would be to see the effects of applying this process to methods with angular features (i.e.\ Angle,
    Interior Angle, Pyramid).
    These methods met the criterion less frequently, and would likely benefit from attempting to reduce the $R$ set before
    deciding to choose another $b$ set.

    \subsection{Identification Step}\label{subsec:identificationStep}
    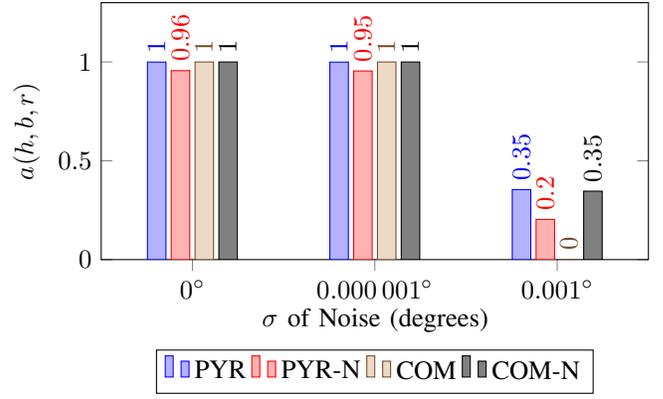
\begin{figure}
        \centering{
        \begin{tikzpicture}
            \begin{axis}[
            ybar,
            width=\linewidth, height=5cm,
            ylabel={$a(h, b, r)$}, ylabel near ticks, ymin=0, ymax=1.3,
            xtick={1, 2, 3}, xticklabels={$\ang{0}$, $\ang{0.000001}$, $\ang{0.001}$},
            xlabel={$\sigma$ of Noise (degrees)}, xmin=0.5, xmax=3.5, xtick pos=left,
            nodes near coords, every node near coord/.append style={rotate=90, anchor=west},
            legend style={at={(0.5,-0.35)}, anchor=north,legend columns=-1},
            bar width=7
            ]
                \addplot coordinates {(1, 0.999416666666667) (2, 0.999222222222222) (3, 0.354333333333333)};
                \addplot coordinates {(1, 0.956083333333329) (2, 0.954444444444449) (3, 0.203166666666665)};
                \addplot coordinates {(1, 1.0) (2, 1.0) (3, 0.0)};
                \addplot coordinates {(1, 1.0) (2, 0.999833333333333) (3, 0.346333333333333)};
                \legend{PYR, PYR-N, COM, COM-N}
            \end{axis}
        \end{tikzpicture}
        \caption{
        Depicts the frequency of correct bijections $a(h, b, r)$ formed with and without the verification step of
        both the Pyramid and Composite Pyramid methods.
        There exists $2{,}000$ runs for each identification method, with a 500 catalog access limit.
        The `-N' suffix indicates the method does not run with the verification step.
        }\label{fig:verify}
        }
    \end{figure}

    \begin{figure*} 
        \centering{
        \begin{subfigure}[b]{0.48\linewidth}
            \begin{tikzpicture}
                \begin{axis}[
                ybar,
                width=\linewidth, height=5cm,
                ylabel={$t \ (\si{ms})$}, ylabel near ticks, ymin=10, ymax=100000,
                xtick={1, 2, 3}, xticklabels={$\ang{0}$, $\ang{0.0001}$, $\ang{0.01}$},
                xlabel={$\rho$}, xmin=0.5, xmax=3.5, xtick pos=left, point meta=rawy,
                nodes near coords, every node near coord/.append style={rotate=90, anchor=west,
                /pgf/number format/.cd,fixed,precision=6},
                legend style={at={(0.5,-0.35)}, anchor=north,legend columns=-1},
                bar width=7, ymode=log, log origin=infty, max space between ticks=20
                ]
                    \addplot coordinates {(1, 1197.43) (2, 1349.07) (3, 6509.00)};
                    \addplot coordinates {(1, 160.68) (2, 161.80) (3, 872.22)};
                    \addplot coordinates {(1, 323.94) (2, 339.23) (3, 392.29)};
                    \addplot coordinates {(1, 324.35) (2, 328.85) (3, 352.24)};
                    \addplot coordinates {(1, 170.79) (2, 171.53) (3, 364.04)};
                    \addplot coordinates {(1, 405.53) (2, 1545.45) (3, 362.30)};
                    \legend{ANG, INT, SPH, PLN, PYR, COM}
                \end{axis}
            \end{tikzpicture}
        \end{subfigure}
        \begin{subfigure}[b]{0.48\linewidth}
            %
            %
            %
            %
            %
            %
            %
            %

            \begin{tikzpicture}
                \begin{axis}[
                width=0.8\linewidth, height=5.3cm,
                ylabel={$a(h, b, r)$}, ymin=0, ymax=1.1,
                xlabel={$\rho$ (degrees)}, xmin=1, xmax=8,
                xtick={1, 2, 3, 4, 5, 6, 7, 8},
                xticklabels={$0$, $10^{-6}$, $10^{-5}$, $10^{-4}$, $10^{-3}$, $10^{-2}$, $10^{-1}$, $1$},
                samples=100, no markers, legend pos=outer north east, enlargelimits=false
                ]
                    \addplot +[domain=1:4, forget plot]{1};
                    \addplot +[forget plot] coordinates {(4, 1) (4, 0.8055384229041530)};
                    \addplot +[domain=4:8]{-1.416887591459587*ln(x) + 2.7697617012853217};
                    \addlegendentry{ANG}

                    \addplot +[domain=1:5.03, forget plot]{1};
                    \addplot +[domain=5.02:8]{-2.3292306442173216*ln(x) + 4.757244753386215};
                    \addlegendentry{INT}

                    \addplot +[domain=1:3, forget plot]{1};
                    \addplot +[forget plot] coordinates {(3, 1) (3, 0.9738787471803161)};
                    \addplot +[domain=3:8]{-1.1317828984553227*ln(x) + 2.217269347527745};
                    \addlegendentry{SPH}

                    \addplot +[domain=1:3, forget plot]{1};
                    \addplot +[domain=3.04:8]{-1.1696544292494204*ln(x) + 2.301996347627258};
                    \addlegendentry{PLN}

                    \addplot +[domain=1:4, forget plot]{1};
                    \addplot +[forget plot] coordinates {(4, 1) (4, 0.8105837347641904)};
                    \addplot +[domain=4:8]{-1.4347255837709008*ln(x) + 2.7995357213002334};
                    \addlegendentry{PYR}

                    \addplot +[domain=1:2, forget plot]{1};
                    \addplot +[forget plot] coordinates {(2, 1) (2, 0.8533960244900531)};
                    \addplot +[domain=2:8]{-0.7510156659772902*ln(x) + 1.3739604159185614};
                    \addlegendentry{COM}
                \end{axis}
            \end{tikzpicture}
        \end{subfigure}
        \caption{
        Both plots represent some statistic about the resulting bijection $h$ produced by each identification method
        given some image with varying Gaussian noise.
        There exist $2{,}000$ runs for each identification method, with a 500 catalog access limit.
        The left plot depicts the average time to obtain $h$, and the right plot depicts the trend line
        $a(h, b, r) = c \cdot \mathit{\ln}\left( \rho \right) + d$.
        }\label{fig:gaussianNoise}
        }
    \end{figure*}

    \subsubsection{How effective are additional verification steps?}
    %
    %
    %
    %
    %
    %

    In~\autoref{fig:verify}, the accuracy of the bijection produced by the Pyramid and Composite Pyramid methods are
    displayed with and without the verification step for varying levels of Gaussian noise.
    Without noise, the Pyramid method without its verification step is 4.33\% less accurate than the Pyramid method with
    verification on average.
    This behavior is consistently seen for Gaussian noise of $\rho\seq\ang{0.000001}$ \& $\rho\seq\ang{0.001}$, and
    can be attributed to the more frequent rejection of incorrect bijections with $R$ sets that have met the criterion.
    In the $\rho\seq\ang{0.001}$ case, there exists a difference of 389.95 accesses between both variations of the Pyramid
    and a 15\% bijection accuracy difference in favor of the method with the verification step.
    Given the null hypotheses that the difference between both variations of the Pyramid method are different for each level
    of noise, $z_0 \seq 15.87, z_{0.000001} \seq 16.04, z_{0.001} \seq 12.14$ (all $p\!<\!0.0001$) is obtained with
    two-tailed two sample $Z$ tests.
    The verification step increases the accuracy of the Pyramid method.

    The response to Gaussian noise for the Composite Pyramid begins at $\rho\seq\ang{0.001}$, with a 34.6\% difference
    between the two variants in favor of the method without the verification step.
    Unlike the verification step in the Pyramid method, this filter appears to be too aggressive for the Composite Pyramid
    method.
    The variant without the verification step has an average of 193.93 catalog accesses at $\rho\seq\ang{0.001}$.
    The Pyramid variant without the verification step only had an average of 8.114 catalog accesses, suggesting that the
    $\abs{R} \seq 1$ criterion and the \Call{DMT}{} process are sufficient enough for rejecting incorrect $r$ sets and
    bijections for the Composite Pyramid method.

    \subsection{End to End}\label{subsec:endToEndEvaluation}
    \subsubsection{Which method is the fastest given no noise?}
    In~\autoref{fig:gaussianNoise}, the left plot depicts the end to end running time of each identification method given
    varying degrees of Gaussian noise.
    In the no noise case, the Angle method is the slowest identification method on average.
    The next slowest method is the Composite Pyramid method, a factor of 2.95 times faster than the Angle method.
    Recall that the Angle method had the fastest query step, but the largest $\abs{R}$.
    On average, it takes 69.85 catalog accesses to obtain a bijection and 68.10 catalog accesses to obtain $r$.
    This suggests that the Angle method's long running time stems from the $\abs{R} \seq 1$ criterion and not the
    \Call{DMT}{} process.

    The fastest method in the no noise case appears to be the Interior Angle method, with the second fastest method running
    10.11\si{ms} slower.
    There exists $0 / 2{,}000$ runs where the Interior Angle method runs above the Pyramid method's average running time
    ($170.79\si{ms}$) and the Interior Angle method has the fastest recorded identification run of 135ms.
    The Interior Angle method is the fastest identification method given no noise.

    \subsubsection{Which method is the fastest given varying levels of Gaussian noise?}
    As Gaussian noise is increased from $\rho\seq\ang{0}$ to $\rho=\ang{0.01}$, the Angle method experiences the largest
    response of $5{,}311.57$ additional $\si{ms}$.
    The next slowest method in the noise of $\rho\seq\ang{0.01}$ case is the Interior Angle method, a factor of 7.46 times
    faster than the Angle method.
    On average, the Angle method takes 399.66 catalog accesses to obtain a bijection and only 36.72 catalog accesses
    to obtain $r$ here.
    In the no noise case, this method's long running time can attributed to the aggressive $R$ criterion.
    Given Gaussian noise, the \Call{DMT}{} process plays a larger role with the Angle method and returns to $b$
    decision process more often.

    The Composite Pyramid method shows an interesting runtime response to this type of noise, running $1{,}139.92\si{ms}$
    longer given $\ang{0.0001}$ of noise from no noise but $1{,}183.15\si{ms}$ shorter from $\ang{0.0001}$ of noise to
    $\ang{0.01}$.
    The Pyramid method is observed to have this same running time response against noise at $\rho\seq\ang{0.001}$
    (not depicted).
    The most probable explanation lies in how far each run travels from the $b$ decision step.
    At $\rho\seq\ang{0.0001}$, the Composite Pyramid has gone through the $\abs{R} \seq 1$ criterion and is likely choosing
    another $b$ set after the verification step.
    At $\rho\seq\ang{0.01}$ the method is not passing the same criterion, avoiding the verification step.

    %
    %
    %
    %
    The fastest method on average given images with the set of Gaussian noise below is the Pyramid method at
    $288.44\si{ms}$ (of 12,000 runs).
    \begin{equation}\label{eq:sigmasTested}
    \rho \in \set{10^{-1}, 10^{-2}, \ldots, 10^{-6}}
    \end{equation}
    The second fastest method given the same noise set is the Planar Triangle method at $341.16\si{ms}$.
    Given the null hypothesis that the difference between both averages is not significant, $z \seq 24.32, p\!<\!0.0001$ is
    found with a two-tailed two sample $Z$ test.
    With the data collected here, the Pyramid method is the fastest method given varying amounts of Gaussian noise.

    \subsubsection{Which method has the slowest growing $h$ accuracy response to increasing noise?}
    The selection of the query $\sigma$ parameters play a significant role in accuracy of each method given images with
    Gaussian noise.
    For methods that query the catalog using on the $\theta$ feature (Angle, Interior Angle, Pyramid), the $\sigma$
    parameter serves as a rough upper bound for the amount of Gaussian noise tolerated.
    When the level of noise is equal to the Angle and Pyramid $\sigma_\theta$ parameter ($\ang{0.0001}$),
    both methods have an average $h$ accuracy of $98.59\!\pm\!1.34\%$.
    When Gaussian noise is increased to $\ang{0.001}$, both methods drop to $47.02\!\pm\!1.58\%$.

    For methods with features that are not angular (Spherical Triangle, Planar Triangle, Composite Pyramid),
    characterizing the effect of Gaussian noise becomes more difficult.
    These methods have the parameters $\sigma_a \seq 10^{-9}$ and $\sigma_\tau \seq 10^{-9}$, showing an initial
    accuracy response to noise at $\ang{0.00001}$.

    Ranking each method based their $h$ accuracy is not particularly insightful here given the heavy dependence on
    $\sigma$ parameters, so instead we analyze the rate of change involved with varying levels of noise.
    The right plot in~\autoref{fig:gaussianNoise} depicts the trend line for all methods where $h$ accuracy is displayed
    against the amount of Gaussian noise.
    It has been observed that the accuracy of each method remains near 100\% until it decreases exponentially to zero.
    As such, each line was fit to the piecewise equation below.
    The $c\cdot \mathit{ln}(\rho) + d$ term was fit using least squares:
    \begin{equation}
        a(f, b, r) =
        \begin{cases}
            0 & \rho < 0 \\
            1 & 0 \leq \rho < \rho^{\star} \\
            c \cdot \mathit{ln}(\rho) + d & \rho \geq \rho^{\star}
        \end{cases}
    \end{equation}
    where $c$ and $d$ are the parameters found with the regression, $a(f, b, r)$ is the accuracy of the bijection, and
    $\rho^{\star}$ is the point where $a(f, b, r)$ is observed to dip below 95\%.
    The accuracy acceleration varies across methods through the value of $c$:
    \begin{equation}
        \frac{d^{2}a(f, b, r)}{d\rho^2} = \frac{-c}{\rho^2}
    \end{equation}
    A larger $c$ suggests that a change in query $\sigma$ or Gaussian noise will not affect the accuracy of the method
    as much as a method with a larger $c$.
    The method with the largest acceleration toward $0\%$ $h$ accuracy is the Interior Angle method ($c \seq -0.15749$).
    The Spherical Triangle method has the slowest growing $h$ accuracy response to increasing noise ($c \seq -0.09266$).

    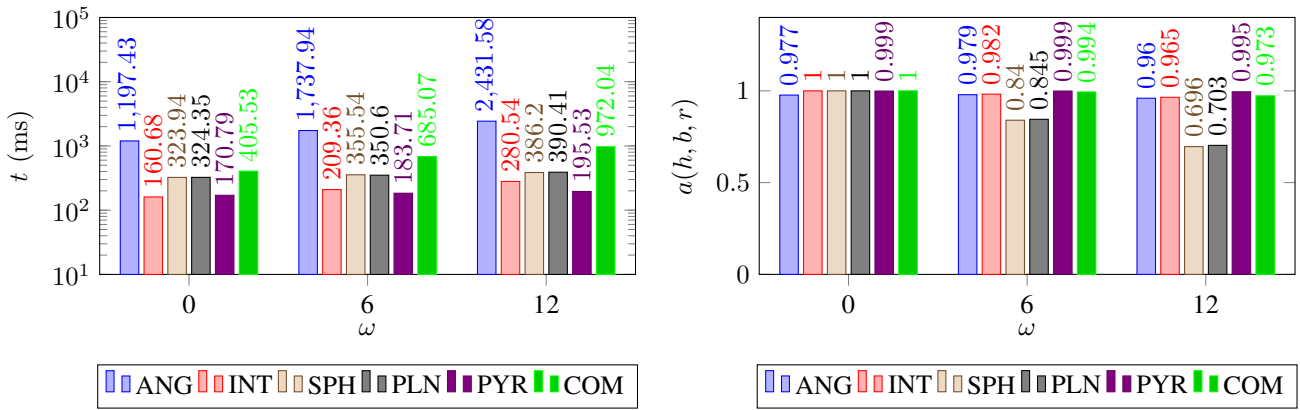
\begin{figure*}
        \centering{
        \begin{subfigure}[b]{0.48\linewidth}
            \begin{tikzpicture}
                \begin{axis}[
                ybar,
                width=\linewidth, height=5cm,
                ylabel={$t \ (\si{ms})$}, ylabel near ticks, ymin=10, ymax=100000,
                xtick={1, 2, 3}, xticklabels={0, 6, 12},
                xlabel={$\omega$}, xmin=0.5, xmax=3.5, xtick pos=left, point meta=rawy,
                nodes near coords, every node near coord/.append style={rotate=90, anchor=west,
                /pgf/number format/.cd,fixed,precision=2},
                legend style={at={(0.5,-0.35)}, anchor=north,legend columns=-1},
                bar width=7, ymode=log, log origin=infty, max space between ticks=20
                ]
                    \addplot coordinates {(1, 1197.43) (2, 1737.94) (3, 2431.58)};
                    \addplot coordinates {(1, 160.68) (2, 209.36) (3, 280.54)};
                    \addplot coordinates {(1, 323.94) (2, 355.54) (3, 386.20)};
                    \addplot coordinates {(1, 324.35) (2, 350.60) (3, 390.41)};
                    \addplot coordinates {(1, 170.79) (2, 183.71) (3, 195.53)};
                    \addplot coordinates {(1, 405.53) (2, 685.07) (3, 972.04)};
                    \legend{ANG, INT, SPH, PLN, PYR, COM}
                \end{axis}
            \end{tikzpicture}
        \end{subfigure}
        \begin{subfigure}[b]{0.48\linewidth}
            \begin{tikzpicture}
                \begin{axis}[
                ybar,
                width=\linewidth, height=5cm,
                ylabel={$a(h, b, r)$}, ylabel near ticks, ymin=0, ymax=1.4,
                xtick={1, 2, 3}, xticklabels={0, 6, 12},
                xlabel={$\omega$}, xmin=0.5, xmax=3.5, xtick pos=left,
                nodes near coords, every node near coord/.append style={rotate=90, anchor=west,
                /pgf/number format/.cd,fixed,precision=4},
                legend style={at={(0.5,-0.35)}, anchor=north,legend columns=-1},
                bar width=7
                ]
                    \addplot coordinates {(1, 0.977) (2, 0.979) (3, 0.960)};
                    \addplot coordinates {(1, 1.0) (2, 0.982) (3, 0.965)};
                    \addplot coordinates {(1, 1.0) (2, 0.840) (3, 0.696)};
                    \addplot coordinates {(1, 1.0) (2, 0.845) (3, 0.703)};
                    \addplot coordinates {(1, 0.999) (2, 0.999) (3, 0.995)};
                    \addplot coordinates {(1, 1.0) (2, 0.994) (3, 0.973)};
                    \legend{ANG, INT, SPH, PLN, PYR, COM}
                \end{axis}
            \end{tikzpicture}
        \end{subfigure}
        \caption{
        Both plots represent some statistic about the resulting bijection $h$ produced by each identification method
        given some image with varying amounts of spikes $\omega$.
        There exist $2{,}000$ runs for each identification method, with a 500 catalog access limit.
        The left plot depicts the average time to obtain $h$, and the right plot depicts the average accuracy of $h$.
        }\label{fig:falseNoise}
        }
    \end{figure*}

    \subsubsection{Which method is the fastest given varying \\ amounts of false stars?}
    In~\autoref{fig:falseNoise}, the plot on the left depicts the end to end running time of each method given varying
    amounts of spikes.
    As the number of spikes increases from 0 to 12, the Angle method again experiences the largest response of
    $1{,}234.15\si{ms}$.
    The next slowest method is the Composite Pyramid method, a factor of 2.50 times faster than the Angle method.
    The difference between the 1st and 2nd slowest methods is 2.98 times less than the Gaussian noise case.
    On average, it takes 114.23 catalog accesses to obtain $h$ and only 54.85 accesses to obtain $r$.
    Relative to the Gaussian noise comparison, \Call{DMT}{} and $\abs{R} \seq 1$ criterion play a more equal role in
    the decision to choose a new $b$ set.

    %
    %
    %
    %
    The fastest method on average given images with varying amounts of spikes is the Pyramid method at $186.22\si{ms}$.
    The images given to each method contained $\omega$ spikes, as defined below:
    \begin{equation}\label{eq:numberOfSpikes}
    \omega \in \set{ 3, 6, 9, 12 }
    \end{equation}
    The second fastest method given the same noise set is the Interior Angle method at $228.37\si{ms}$.
    Given the null hypothesis that the difference between both averages is not significant, $z \seq 28.47, p\!<\!0.0001$ is
    found with a two-tailed two sample $Z$ test.
    The Pyramid method is the fastest method given varying amounts of spikes.
    The process for choosing distinct image star sets is shown to be effective in finding a bijection that meets the Pyramid
    criteria the fastest.

    %
    %
    %
    Each method exhibits a linear increase to runtime as additional spikes are added.
    To characterize how each method's runtime grows with increasing false stars, each method's runtime was fit to a linear
    equation using least squares:
    \begin{equation}
        t = c\cdot\omega + d
    \end{equation}
    where $c$ and $d$ are the parameters found with the regression and $t$ is the end to end running time of the method.
    A smaller $\abs{c}$ suggests that the number of spikes will affect the end to end runtime than that of a method with a
    larger $\abs{c}$.
    The method with the largest $\abs{c}$ is the Angle method with $c \seq -414.559$.
    The method with the smallest $\abs{c}$ term is the Pyramid method with $c \seq -6.766$.
    The Pyramid method is the fastest given varying amounts of false stars, having a runtime that is also the least
    responsive to increasing spikes.

    \subsubsection{Which method is the most accurate given varying amounts of false stars?}
    %
    %
    %
    %
    In~\autoref{fig:falseNoise}, the plot on the right depicts the average accuracy of each bijection given varying amounts
    of spikes.
    As the number of false stars is increased from $\omega \seq 0$ to $\omega \seq 12$, the methods that experience the largest
    $h$ accuracy response are the Spherical Triangle method ($30.42\%$ average decrease) and the Planar Triangle method
    ($29.68\%$ average decrease).
    The average accuracy of the $r$ selection is a few percent less than the average accuracy of $h$ here
    ($0.53\!\pm\!1.78\%$ for both methods).
    Given the null hypothesis that the difference between the accuracy of the $h$ bijection and the accuracy of the $r$
    selection is not significant, $z \seq 0.37, p \seq 0.71$ was found with a two-tailed two sample $Z$ test.
    There does not exist enough data to reject this hypothesis with $\alpha \seq 0.01$.
    This suggests that the \Call{DMT}{} process is neither helpful or detrimental to the end to end accuracy of these
    methods.

    Ruling out the \Call{DMT}{} process, the most likely source of error for the triangle methods is their decision of
    different $b$ sets.
    If a false star exists as $b_1$ in $b$, the triangle methods will have to iterate through $n^2$ combinations and $n - 3$
    pivots at most to choose another star that is not the spike.
    The Angle method only has to wait $n$ additional combinations at most if a false star exists in $b$.
    The Interior Angle method is able to get around the spike persistence problem by choosing $b$ sets based on their
    $\theta$ proximity to the central star $b_c$.
    The Pyramid and Composite Pyramid methods have their $b$ decision process designed for this situation, increasing the
    average turnover of all stars in the $b$ set.

    The Pyramid method has the most accurate $h$ on average given images with $\omega$ in~\autoref{eq:numberOfSpikes}
    at $99.84\!\pm\!3.53\%$.
    The second most accurate method is the Composite Pyramid method at $99.19\!\pm\!8.95\%$.
    Given the null hypothesis that the difference between the $h$ accuracies of both methods is not significant,
    $z \seq 3.02, p \seq 0.003$ with a two-tailed two sample $Z$ test.
    At $\alpha \seq 0.01$, our hypothesis does not hold true.
    The Pyramid method is the most accurate under varying amounts of spikes.

    \section{Conclusion}\label{sec:conclusion}
    In this paper, we discussed six star identification methods and their strengths and weaknesses.
    A unified identification framework was created to describe all methods for fair analysis.
    Portions that were interchangeable amongst all methods such as database access and centroid determination were
    normalized or removed to focus on the star identification aspect itself.
    To control the severity of our error, artificial images were generated.

    The Angle method is the simplest of the six and has the fastest query step, but its runtime is heavily impacted by
    the $\abs{R} = 1$ criterion and \Call{DMT}{} process.
    The Interior Angle method is the fastest running method under no noise, but its accuracy is the most sensitive to
    varying Gaussian noise and the slowest query step.
    The Spherical Triangle method's accuracy is the least sensitive to varying Gaussian noise, but is the most sensitive
    to varying amounts of false stars.
    The Planar Triangle is on average faster than the Spherical Triangle method, but is also very sensitive to varying
    amounts of false stars.
    The Pyramid method is the fastest method given varying amounts of Gaussian noise \& false stars and is also the most
    accurate given varying amounts of spikes, but is not able to achieve $100\%$ average accuracy due to its query step.
    The Composite Pyramid method does not suffer from this inaccuracy problem, but does not achieve the same consistent
    performance of the Pyramid or the triangle methods due to the number of filters implemented.

    Overall, the Pyramid method handles both Gaussian noise and false stars the best in a reasonable amount of time.

    \section{Acknowledgements}\label{sec:acknowledgements}
    We would like to thank Dr. Miguel Nunes, Eric Pilger, and Yosef Ben Gershom from the Hawaii Space Flight Laboratory
    for providing input toward the creation of software for a first generation star tracker.

    \balance

    \bibliographystyle{abbrv}

\end{document}